\documentclass{aa}

\usepackage{graphicx}
\usepackage{txfonts}
\usepackage{natbib}
\bibpunct{(}{)}{;}{a}{}{,} 
\renewcommand{\sun}{\odot}

\begin{document} 

   \title{Submillimetre water masers at 437, 439, 471, and 474 GHz towards evolved stars}

   \subtitle{APEX observations and radiative transfer modelling}

   \author{
          P. Bergman\inst{1}
          \and
          E. M. L. Humphreys\inst{2,3}
          }
        \institute{
          Department of Space, Earth and Environment, Chalmers University of Technology, Onsala Space Observatory, 43992 Onsala, Sweden\\
          \email{per.bergman@chalmers.se}
         \and
           European Southern Observatory, Karl-Schwarzschild-Str. 2, 85748 Garching, Germany\\
           \email{ehumphre@eso.org}
           \and
           Joint ALMA Observatory, Av. Alonso de Cordova 3107, Vitacura, Santiago, Chile\\
        }

   \date{Received 19 February 2020 / Accepted 28 March 2020}

 
  \abstract
   {}
   {Here we aim to characterise submillimetre water masers at 437, 439, 471, and 474~GHz
   towards a sample of evolved stars.}
   {We used the Atacama Pathfinder Experiment (APEX$^{1}$) to observe submillimetre water transitions
    and the CO (4-3) line towards 11 evolved stars. The sample included semi-regular and Mira variables, plus a red
    supergiant star. We performed radiative transfer modelling for the water masers. We also used the CO observations
    to determine mass loss rates for the stars.}
   {From the sample of 11 evolved stars, 7 display one or more of the masers at 437, 439, 471, and 474~GHz.
   We therefore find that these masers are common in evolved star circumstellar envelopes.
   The fact that the maser lines are detected near the stellar velocity
   indicates that they are likely to originate from the inner circumstellar envelopes
   of our targets. We tentatively link the presence of masers to the degree of variability of the target star, that is,
   masers are more likely to be present in Mira variables than in semi-regular variables.
   We suggest that this indicates the importance of strong shocks in creating the necessary conditions for the masers.
   Typically, the 437~GHz line is the strongest maser line observed among those studied here. We cannot reproduce the
   above finding in our radiative transfer models.
   In general, we find that maser emission is very sensitive to dust temperature in the
   lines studied here.
   To produce strong maser emission, the dust temperature must be significantly lower than the gas kinetic
   temperature.  In addition to running grids of models in order to determine the optimum physical conditions for
   strong masers in these lines, we performed smooth wind modelling for which we cannot reproduce the observed line
   shapes.
   This also suggests that the masers must originate predominantly from the inner envelopes.
    }
   {}
   
    \keywords{stars: AGB and post-AGB 
                -- masers -- submillimeter: stars  
                -- stars: winds, outflows
               }
               
               \titlerunning{Water masers at 437, 439, 471, and 474 GHz}
               \authorrunning{Bergman \& Humphreys}

   \maketitle

%

\section{Introduction}

\footnotetext[1]{
Based on observations with the Atacama Pathfinder EXperiment (APEX) telescope under programme IDs 091.F-9329, 093.F-9315, and 095.F-9313.
APEX  is a collaboration between the Max Planck Institute for Radio Astronomy, the European Southern Observatory, and the Onsala Space 
Observatory. Swedish observations on APEX are supported through Swedish Research Council grant number 2017$-$00648.
}

Single-dish telescopes and radio interferometers have been used to perform detailed studies of circumstellar water masers at 22 GHz towards
evolved stars. 
Observations show that 22 GHz masers typically probe regions in which stellar wind formation 
and acceleration is taking place \citep[e.g.][]{Bains2003,Richards2012}, and that they are common towards oxygen-rich evolved stars. 
Surveys at 183, 321, 325, and 658 GHz \citep[][]{Menten1991,Menten1995,Yates1995,Yates1996,Gonzalez1998,Hunter2007,Humphreys2017,Baudry2018}
indicate that millimetre and submillimetre H$_{2}$O masers are also common in circumstellar envelopes (CSEs) of evolved stars, namely
asymptotic giant branch (AGB) stars and red supergiants.
Submillimetre water maser observations may therefore help in the determination 
of quantities needed to understand the AGB mass-loss process such as gas temperature and density, kinematics, and magnetic field estimation.
The first subarcsecond imaging of  submillimetre water masers at 321, 325, and 658 GHz towards an evolved star, the red supergiant VY~CMa,
was performed by \citet{Richards2014}.

The water maser frequencies selected for observation here are those at 437, 439, 471, and 474 GHz.
To date, these  maser lines have not been widely studied.
Maser emission in the 439 and 471 GHz transitions was discovered
by \citet{Melnick1993} using the Caltech Submillimeter Observatory towards three
star-forming regions G34.3-0.2, W49N, and Sgr~B2, and an evolved star, the AGB star U~Her. The 437~GHz line
was additionally detected towards U~Her, but not towards the star-forming regions. The 474 GHz line was detected towards
VY~CMa and the AGB star W~Hya using the APEX FLASH receiver \citep{Menten2008}.
For extragalactic objects, \citet{Humphreys2005} made a tentative detection of the 439 GHz maser towards the active
galactic nuclei (AGN) and starburst galaxy NGC 3079 using the James Clerk Maxwell Telescope.
New observations therefore constitute pathfinder observations for evolved stars and AGNs using
the Atacama Large Millimeter/submillimeter Array Band~8 receiver.

\begin{table}
\begin{center}
\caption{Observed lines}
\label{linelist}
\begin{tabular}{lcccr}
\hline
\hline
Molecule & Frequency &  Transition          & {\it Ortho}/  & $E_u$  \\
         & (GHz)     &  $J_{Ka,Kc}$         & {\it Para}    &  (K)   \\
\hline
H$_2$O & 437.3467    & ${7}_{5,3}-{6}_{6,0}$ &  {\it p}      & 1504   \\
       & 439.1508    & ${6}_{4,3}-{5}_{5,0}$ &  {\it o}      & 1068   \\
       & 470.8889    & ${6}_{4,2}-{5}_{5,1}$ &  {\it p}      & 1042   \\
       & 474.6891    & ${5}_{3,3}-{4}_{4,0}$ &  {\it p}      &  702   \\
\hline
CO     & 461.0407    &  $J=4-3$             & ...           &   55   \\
\hline
$^{29}$SiO & 471.5256 & $v=0$ $J=11-10$     & ...           &  136   \\
SiO        & 474.1850 & $v=1$ $J=11-10$     & ...           & 1906   \\
SO         & 471.5374 & $N_J=11_{10}-10_9$  & ...           &  143   \\
\hline
\end{tabular}
\end{center}
\end{table}

\begin{figure}
\resizebox{\hsize}{!}{ \includegraphics[]{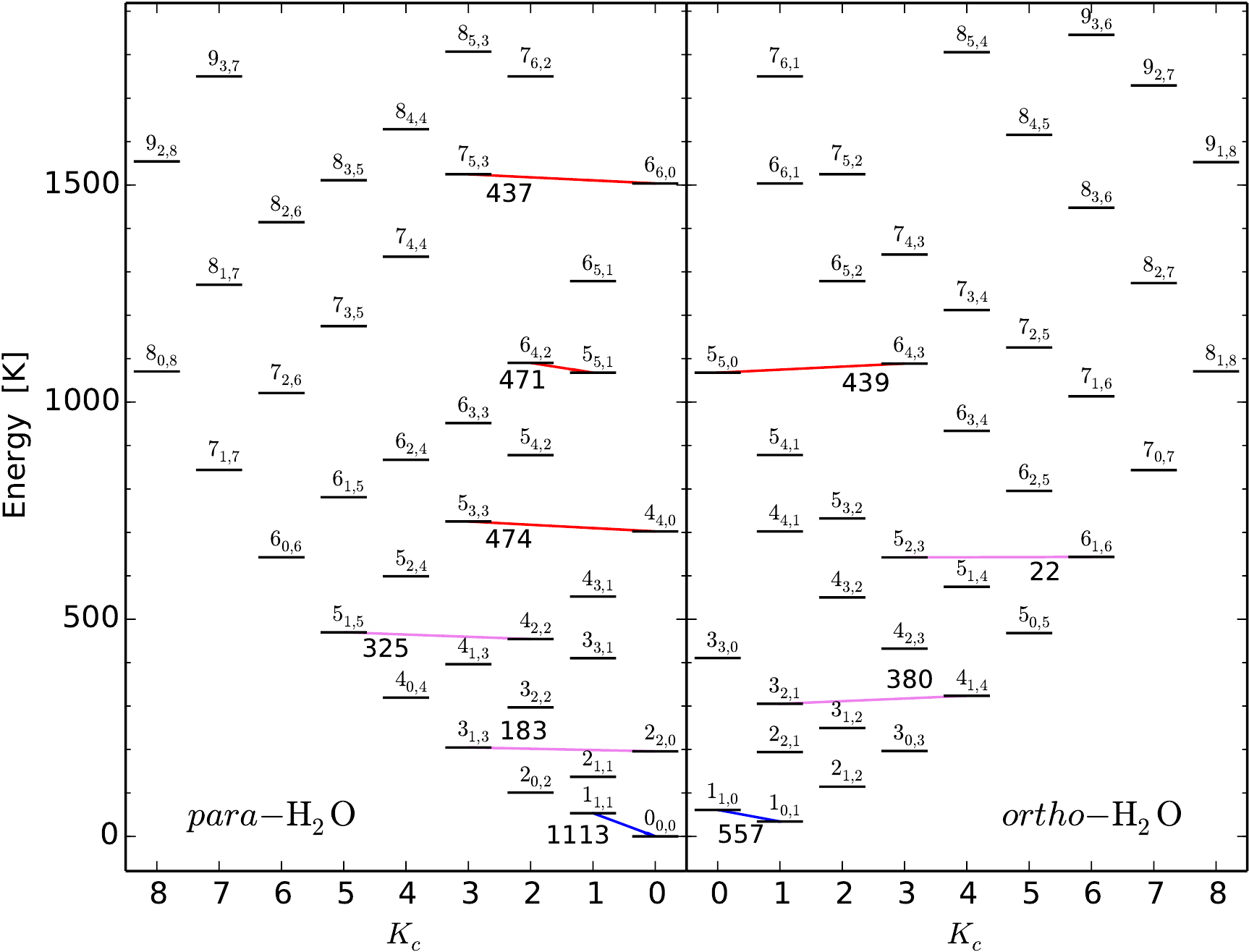} }
     \caption{Water energy level diagram for the lower portion of the vibrational ground state. As indicated,
     {\it para} levels are shown to the left and {\it ortho} levels are shown to the right. The state energy is given in Kelvin and is shown as a function of
     quantum number $K_{c}$. Near each level, the quantum numbers $J_{K_a,K_c}$ are listed. The 400 GHz transitions are shown as red lines, connecting upper and lower levels, with the frequency in gigahertz indicated. Some other relevant masing or thermal transitions are also included.}
               \label{energylevs}       
\end{figure}

\begin{table*}
          \caption{\label{sample} Stellar sample and water maser detections}
\centering
   \begin{tabular}{lcccccccccc}
   \hline\hline
Star     & \multicolumn{3}{c}{Magnitude\tablefoottext{a}} &  $L_\ast$   & $T_\ast$ & Distance\tablefoottext{b} &    \multicolumn{4}{c}{Water masers detected in this work} \\
         &  Type & variation                &  Period     &  (L$_\sun$) & (K)      & (pc)                      & 437            & 439           & 471            & 474 \\
         &       & range  ($\Delta$V)       &    (days)   &             &          &                           & GHz            & GHz           & GHz            & GHz \\
\hline
VX Sgr   & RSG  & 7.5  & 732 &   $1.0\times 10^{5}$ & 3500 & 1600 &   Y    &     Y     &     Y    &   Y  \\
 U Her   & Mira & 7.0  & 404 &   $4.4\times 10^{3}$ & 2700 &  266 &   Y    &     Y     &     Y    &   Y  \\
RR Aql   & Mira & 6.7  & 395 &   $7.3\times 10^{3}$ & 2500 &  633 &   Y    &     N     &     N    &   N  \\
W Hya    & Mira & 4.0  & 390 &   $4.5\times 10^{3}$ & 3100 &  104 &   Y    &     Y     &     Y    &   Y  \\
R Hya    & Mira & 7.4  & 380 &   $7.4\times 10^{3}$ & 2100 &  124 &   N    &     N     &     N    &   N  \\
RS Vir   & Mira & 7.6  & 354 &   $4.4\times 10^{3}$ & 2900 &  610 &   Y    &     Y     &    (Y)   &   Y  \\
R Leo    & Mira & 6.9  & 310 &   $2.5\times 10^{3}$ & 2000 &   95 &   N    &     N     &     N    &   N  \\
R Aql    & Mira & 6.5  & 270 &   $4.4\times 10^{3}$ & 2700 &  422 &   Y    &     Y     &     Y    &   N  \\
R Dor    & SRb  & 1.5  & 172 &   $6.5\times 10^{3}$ & 2400 &   59 &   N    &     N     &    (Y)   &   N  \\
R Crt    & SRb  & ...  & 160 &   $2.1\times 10^{4}$ & 2800 &  261 &   N    &     N     &     N    &   N  \\
RT Vir   & SRb  & 1.6  & 158 &   $4.0\times 10^{3}$ & 2800 &  226 &   N    &     N     &     N    &   N  \\
 \hline
           \end{tabular}
\tablefoot{
\tablefoottext{a}{Information from the AAVSO.} 
\tablefoottext{b}{The distance references are:
 \citet[][VX Sgr]{Chen2007},
 \citet[][RR Aql and U Her]{Vlemmings2007},
 \citet[][RS Vir]{Whitelock2008},
 \citet[][R Leo]{Haniff1995},
 \citet[][R Dor]{Knapp2003},
 \citet[][R Crt, R Aql, R Hya, and W Hya]{vanLeeuwen2007},
 \citet[][RT Vir]{Zhang2017}. }
}

\end{table*}

The  437, 439, 471, and 474 GHz lines are potentially of high interest for the investigation
of CSEs if common and strong. Whereas masers at 22, 183, and 325 GHz originate from the water
`back-bone' levels, masers at 437, 439, 471, and 474 GHz originate from the 
so-called water `transposed back-bone' (Fig.~\ref{energylevs}). While it is generally
accepted that the back-bone water masers are predominantly collisionally pumped followed
by radiative decay, this is not necessarily the case for the transposed back-bone masers.
\citet{Yates1997} find a significant radiative pumping component for these masers that
makes them stronger in the presence of warm dust  ($T_{\mathrm{dust}} > 250\,\mathrm{K}$).  The
evolved star water maser model of \citet{Gray2016} predicts significant
inversions for the maser lines at 439 and 474 GHz, but not those at 437 and
471 GHz. 

Here, we report the results of observations of a pilot survey of submillimetre
water masers and CO (4-3) toward a sample of 11 evolved stars using APEX \citep{Gusten2006}. 
The source selection criteria selected stars that have previous detections of water masers at 22 GHz, and 
most of the sample have also been 
searched for 321 and 325 GHz maser emission by \citet{Yates1995}. We observed some targets 
at multiple epochs to begin determining variability characteristics.
We also attempt to understand the physical conditions needed for strong maser
emission at 437, 439, 471, and 474 GHz using water maser radiative transfer modelling.

\section{Observations}

Observations were made using the Swedish Heterodyne Facility Instrument (SHFI) APEX-3
receiver \citep{Vassilev2008,Belitsky2006} between May 8 2013 and May 20 2015. SHFI APEX-3
was a double-side-band (DSB) heterodyne receiver covering the frequency range 385-500 GHz.
The rest frequencies of the target lines (Table~\ref{linelist})
were covered by two tunings, one centred near 439 GHz and the other at 472.8 GHz. For the latter tuning, the CO 4-3 line emerging from the image band appears
in between the two water lines. The two tunings were observed on the same night towards any
given target (Table~\ref{sample}) such that the observations are quasi-simultaneous. The high-frequency tuning
also includes lines from SO and SiO; see Table~\ref{linelist}.

The instantaneous back-end bandwidth for each tuning was 4 GHz, as made up from two partly overlapping 2.5 GHz Fast Fourier
Transform Spectrometers (FFTS) of 32768 channels \citep{Klein2012}.
Observations were made in beam-switching mode with an azimuthal throw of 100 arcsec and with a
switching frequency of 0.5 Hz to reduce the impact of atmospheric fluctuations.

Weather conditions  were always good during our observations. In 2013 the precipitable
water vapour column was around 0.3~mm while in 2014 and 2015 it was typically in
the range 0.3--0.45~mm. Pointing observations in the CO 4-3
line were made towards the science targets themselves providing the CO line was
strong enough; otherwise nearby carbon-rich CSEs were used. The APEX full half-power beam width at 460 GHz
is 13 arcsec.

Between March 20 and June 13, 2014, the APEX-3 receiver response to the calibration loads was not linear and for a
subset of our observations additional re-calibration factors had to be
applied\footnote{\url{http://www.apex-telescope.org/heterodyne/shfi/calibration/calfactor/}} to the data.
Also, the normal calibration factors for the DSB receiver APEX-3 are those given at the centre of each FFTS 2.5 GHz segment for
the signal band. In the APEX-3 frequency band, the atmospheric transmission can vary substantially over our bands.
As a result, spectral lines coming from the image band, like CO(4-3) in our case, have to be recalibrated due
to the different atmospheric optical depths at signal and image frequencies. This latter recalibration was done for
all sources and all epochs. The main beam efficiency for APEX-3
is $\eta_\mathrm{mb} = 0.60$ and the adopted Jansky-to-Kelvin conversion factor
was $48\,\mathrm{Jy/K}$.\footnote{\url{http://www.apex-telescope.org/telescope/efficiency/index.php.old}}
The typical calibration uncertainty is 10-15\%.

\section{Results}

\begin{figure}
\resizebox{\hsize}{!}{ \includegraphics[width=10cm]{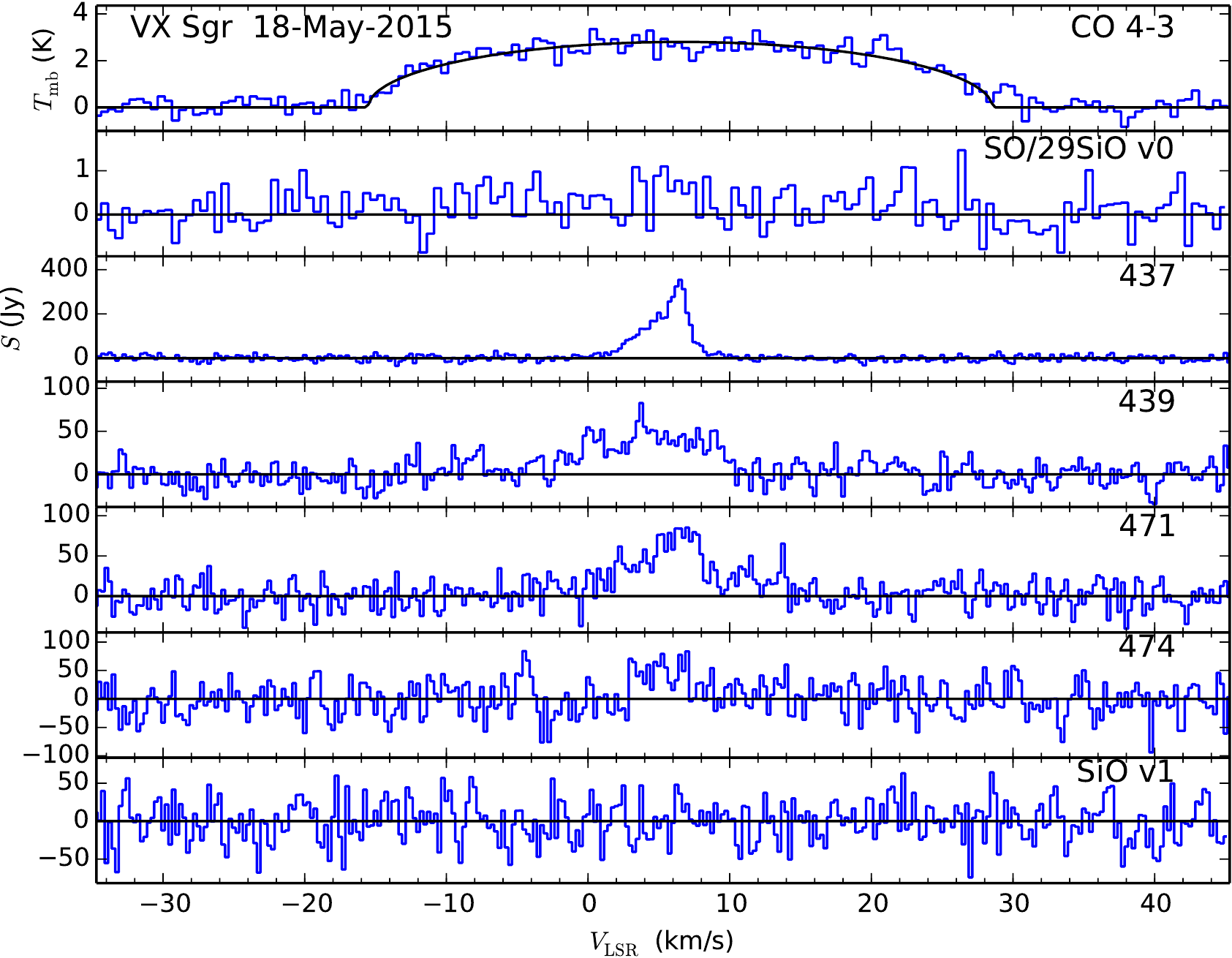} }
\caption{APEX-3 observations towards VX Sgr. The observations of CO, SO, and $^{29}$SiO are reported in
     main beam temperature using a velocity resolution of $0.5\,\mathrm{km\, s^{-1}}$. Observations of
     the water lines and SiO $v=1$ are reported in flux density and are shown with a spectral resolution
     of $0.25\,\mathrm{km\, s^{-1}}$.}
               \label{vxsgr}       
\end{figure}

\begin{figure*}
   \centering
    \includegraphics[width=17cm]{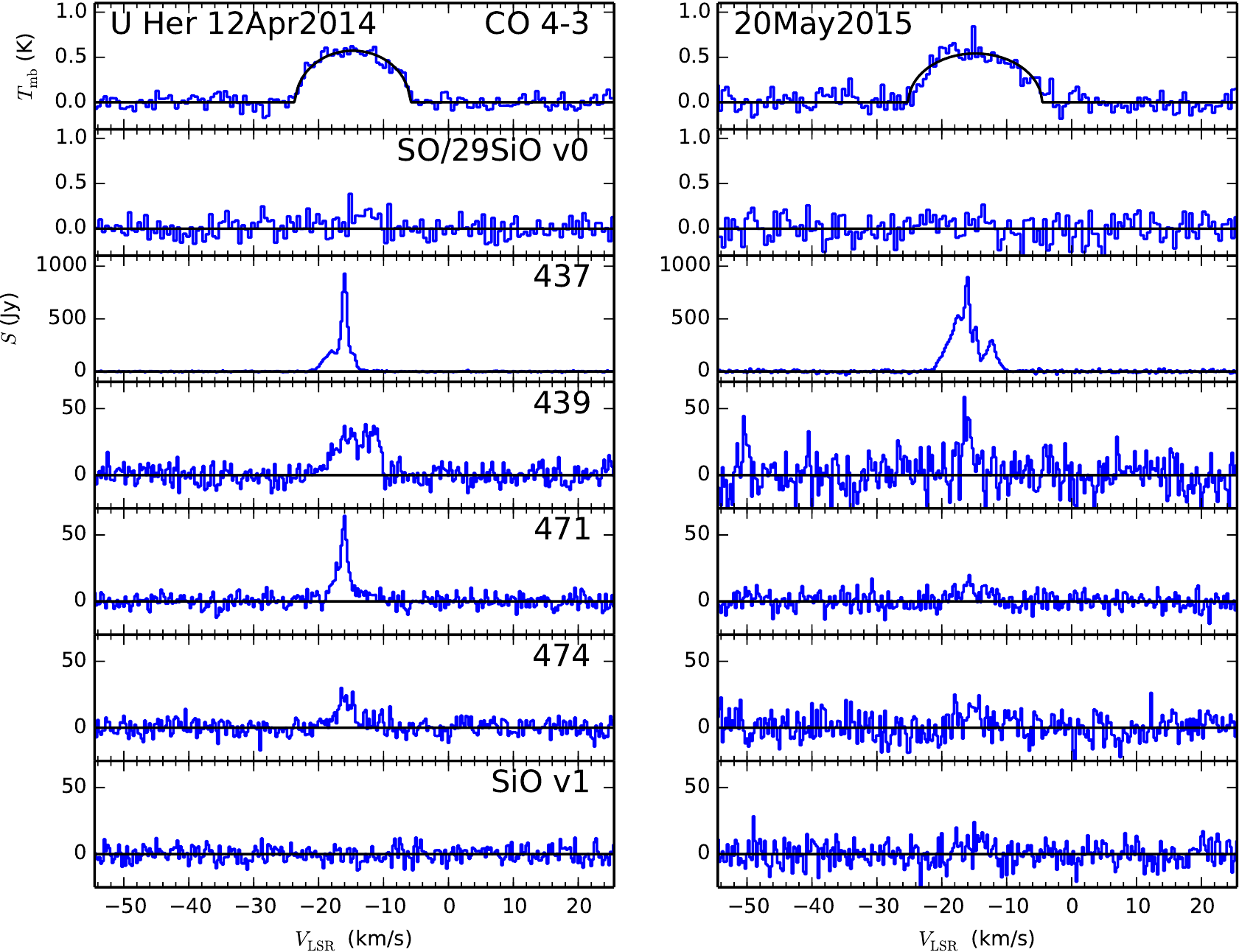}
     \caption{As for Fig.~\ref{vxsgr} but for U Her.}
               \label{uher}       
\end{figure*}

\begin{figure}
 \resizebox{\hsize}{!}{  \includegraphics[width=10cm]{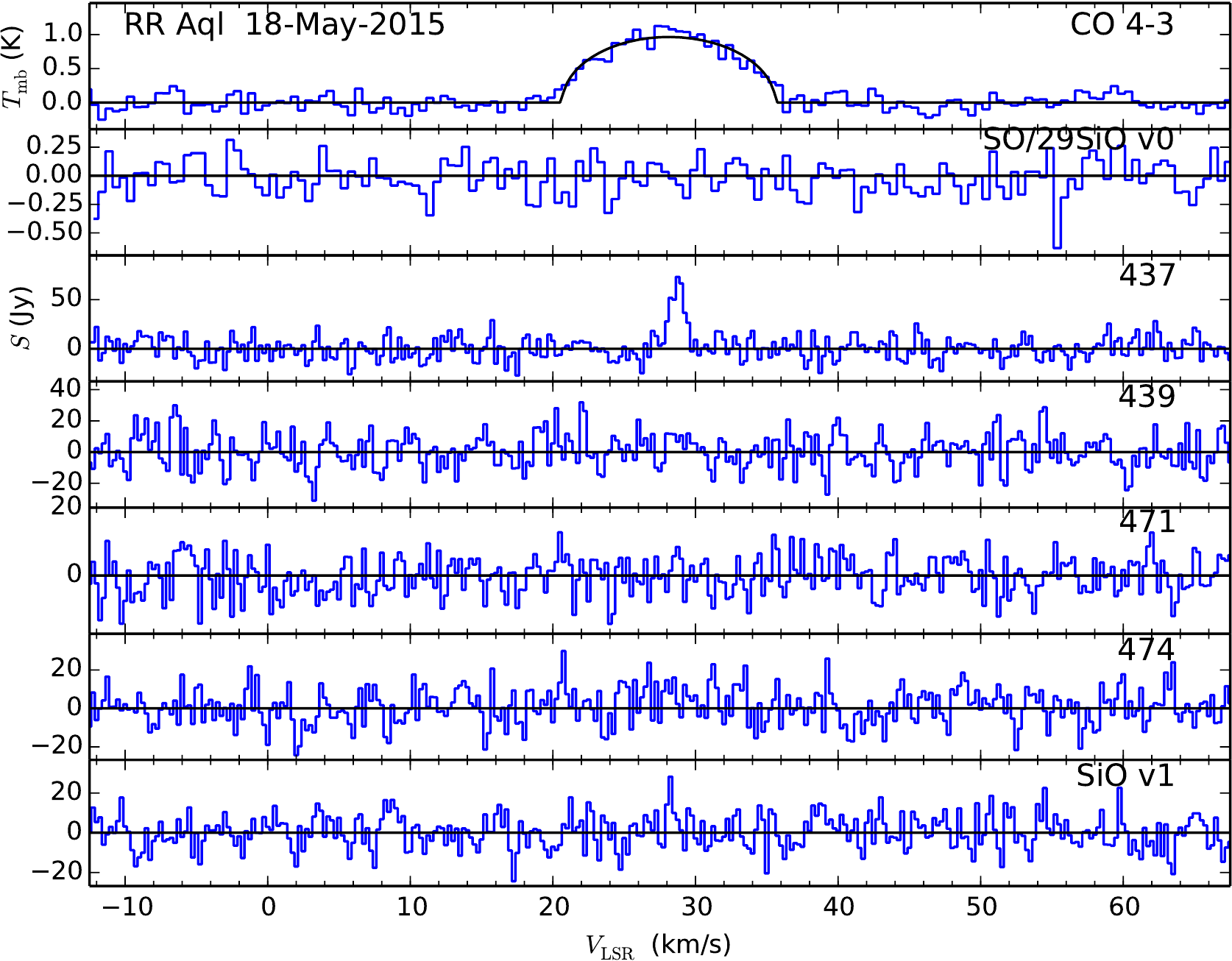} }
     \caption{As for Fig.~\ref{vxsgr} but for RR Aql.}
               \label{rraql}       
\end{figure}

\begin{figure*}
  \centering
  \includegraphics[width=17cm]{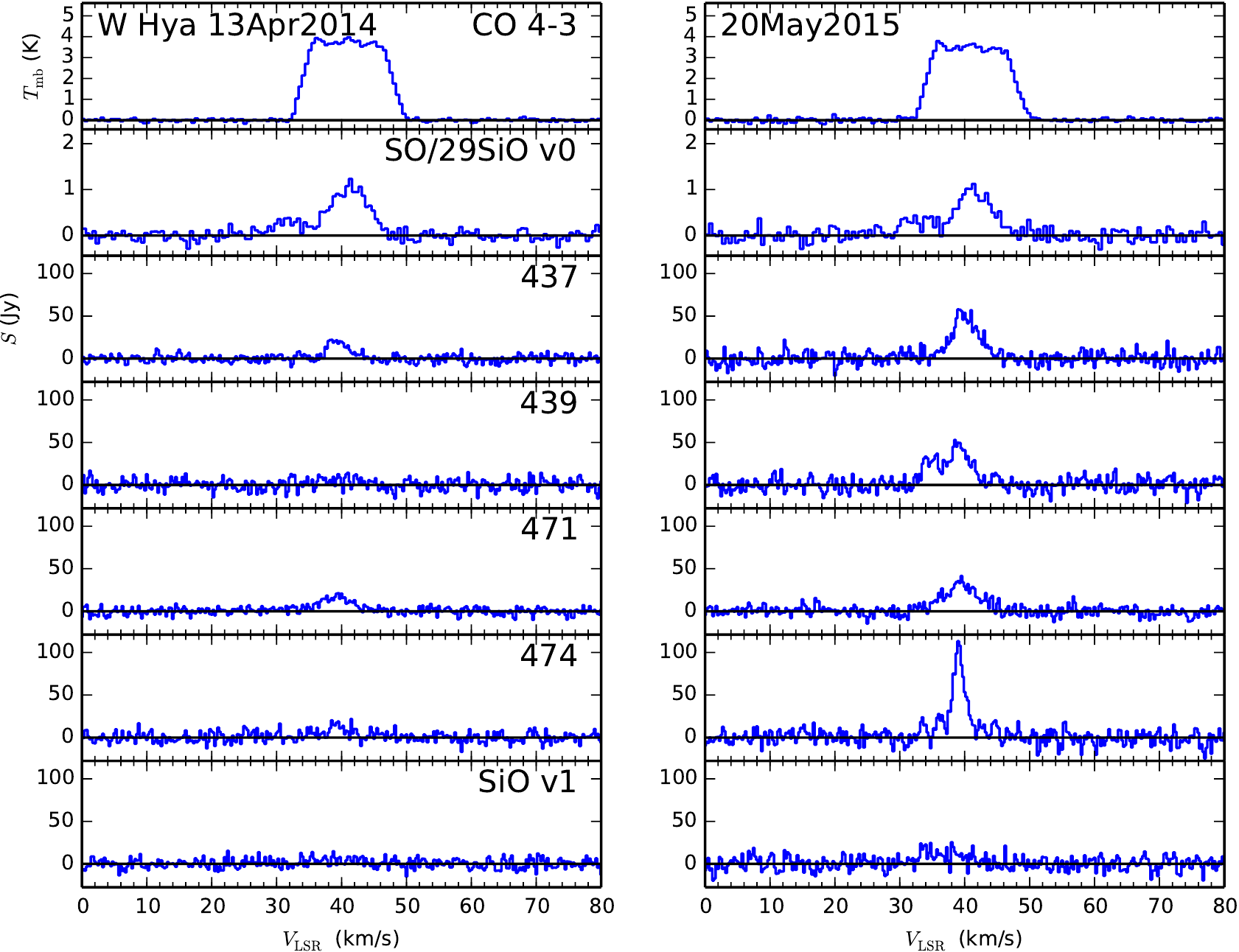}
     \caption{As for Fig.~\ref{vxsgr} but for W Hya.}
               \label{whya}       
\end{figure*}

\begin{figure}
 \resizebox{\hsize}{!}{ \includegraphics[]{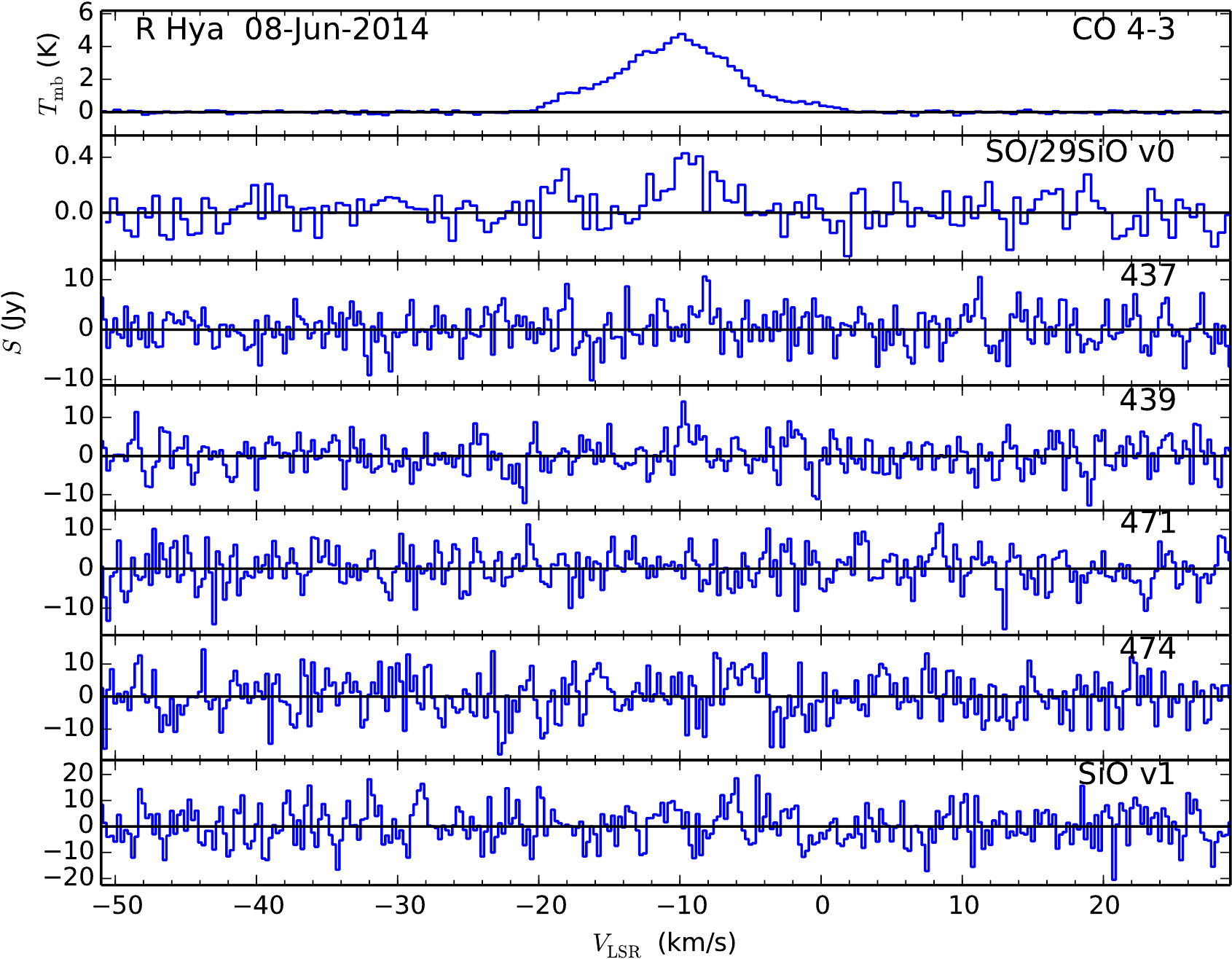} }
     \caption{As for Fig.~\ref{vxsgr} but for R Hya.}
               \label{rhya}       
\end{figure}

\begin{figure*}
   \centering
    \includegraphics[width=17cm]{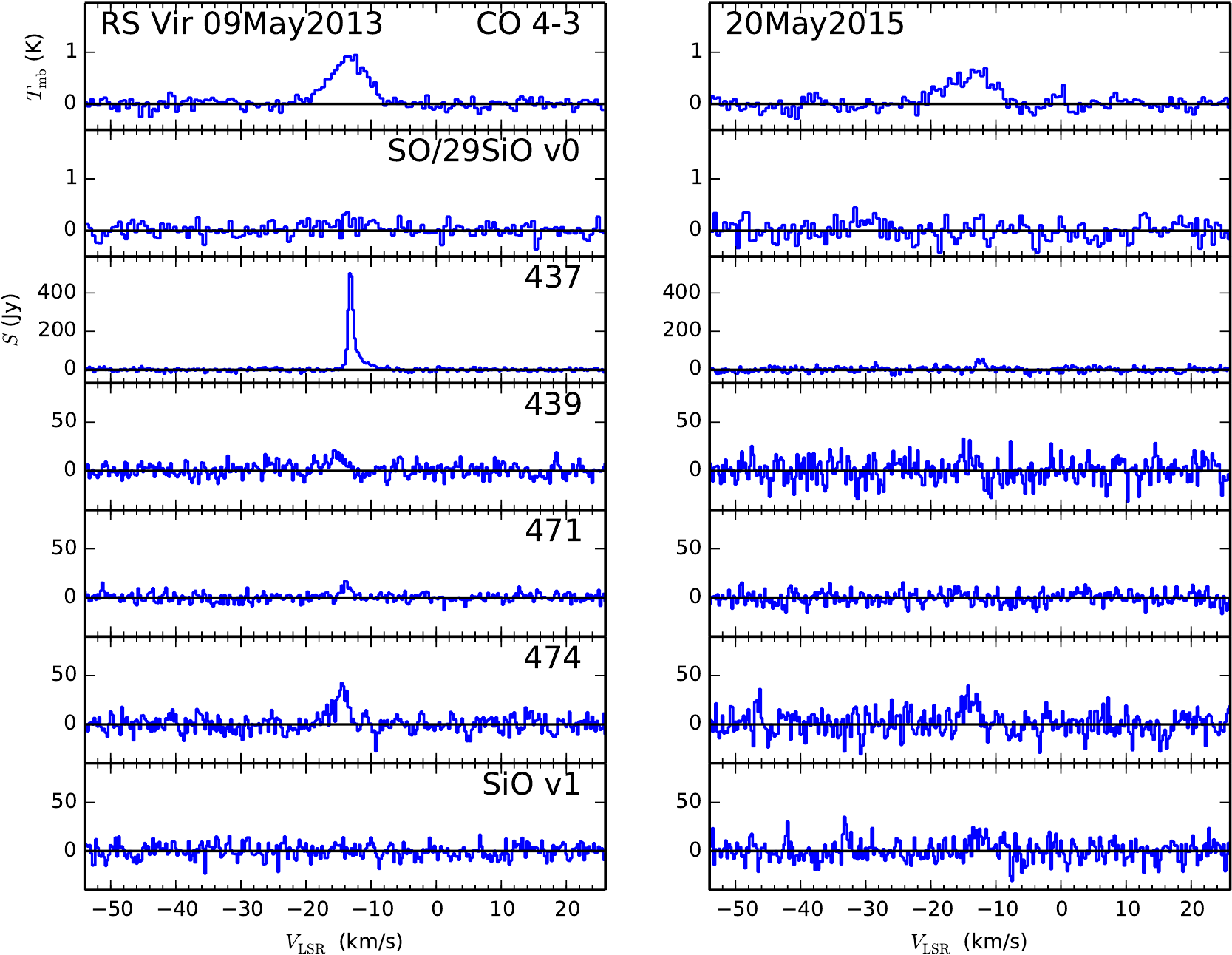}
     \caption{As for Fig.~\ref{vxsgr} but for RS Vir.}
               \label{rsvir}       
\end{figure*}

\begin{figure}
\resizebox{\hsize}{!}{ \includegraphics[]{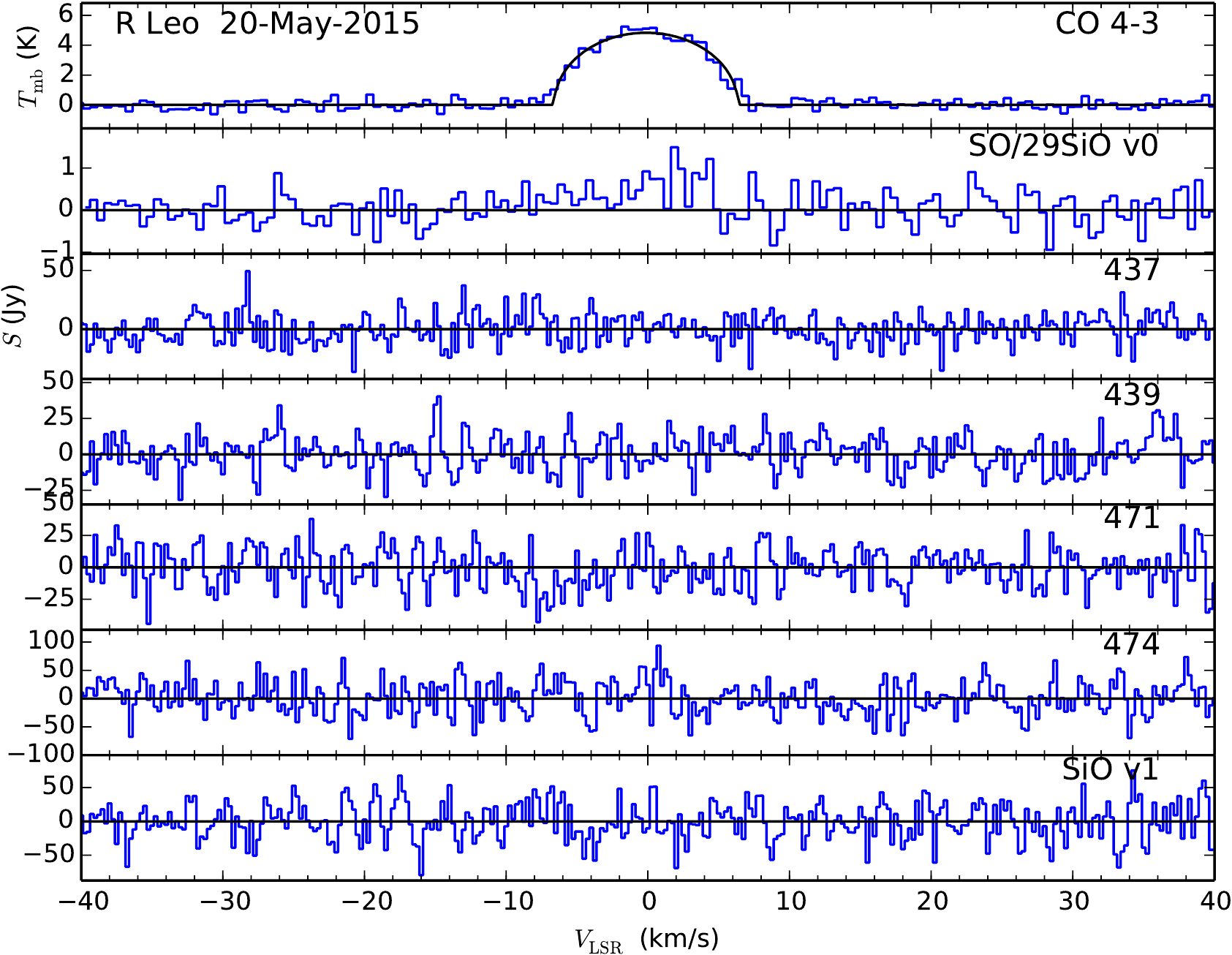} }
     \caption{As for Fig.~\ref{vxsgr} but for R Leo.}
               \label{rleo}       
\end{figure}

\begin{figure*}
   \centering
    \includegraphics[width=17cm]{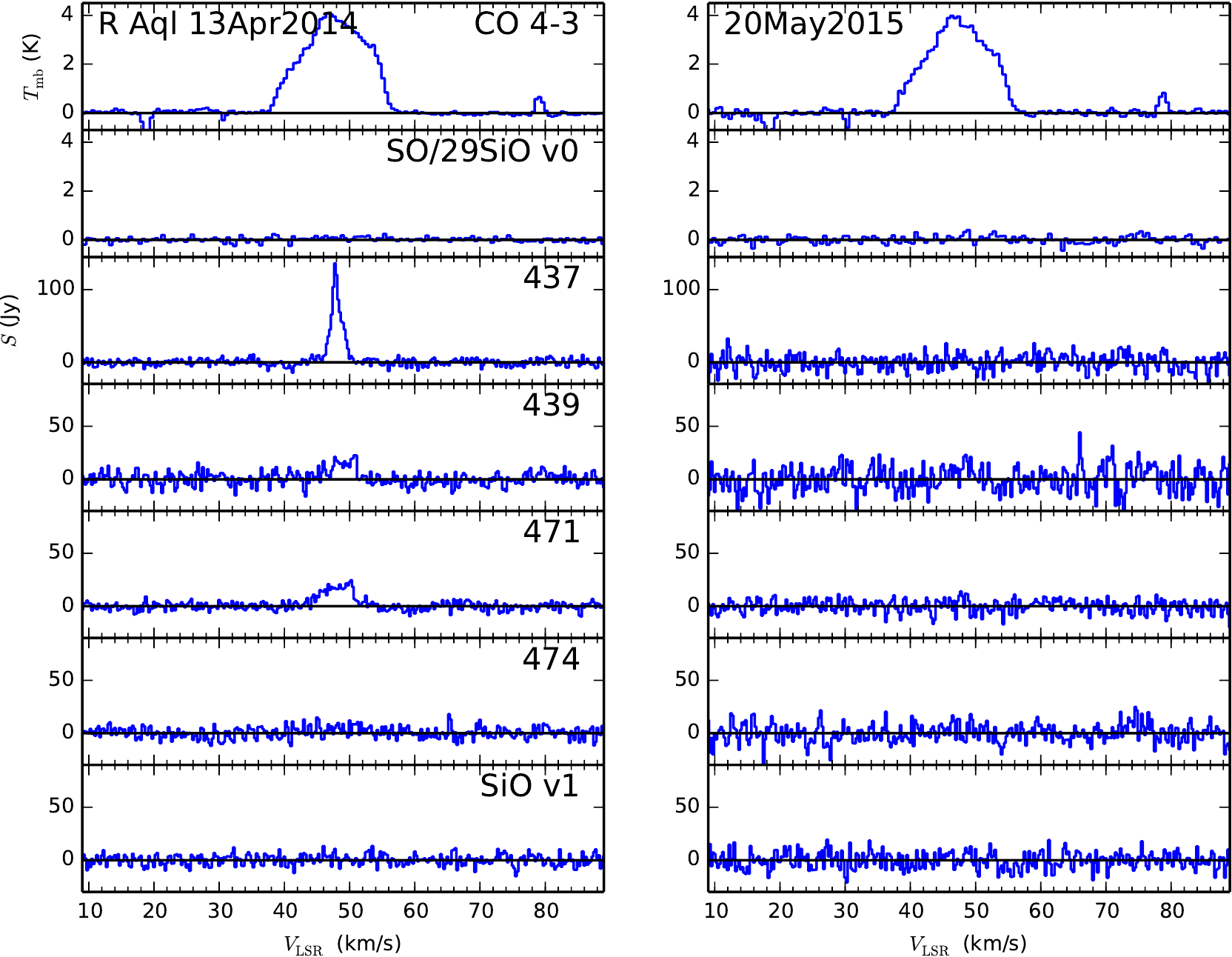}
     \caption{APEX-3 observations towards R Aql. The CO 4-3 features at +20, +30, and +80 $\mathrm{km\, s^{-1}}$ are due to ISM contamination.}
               \label{raql}       
\end{figure*}

\begin{figure}
 \resizebox{\hsize}{!}{ \includegraphics{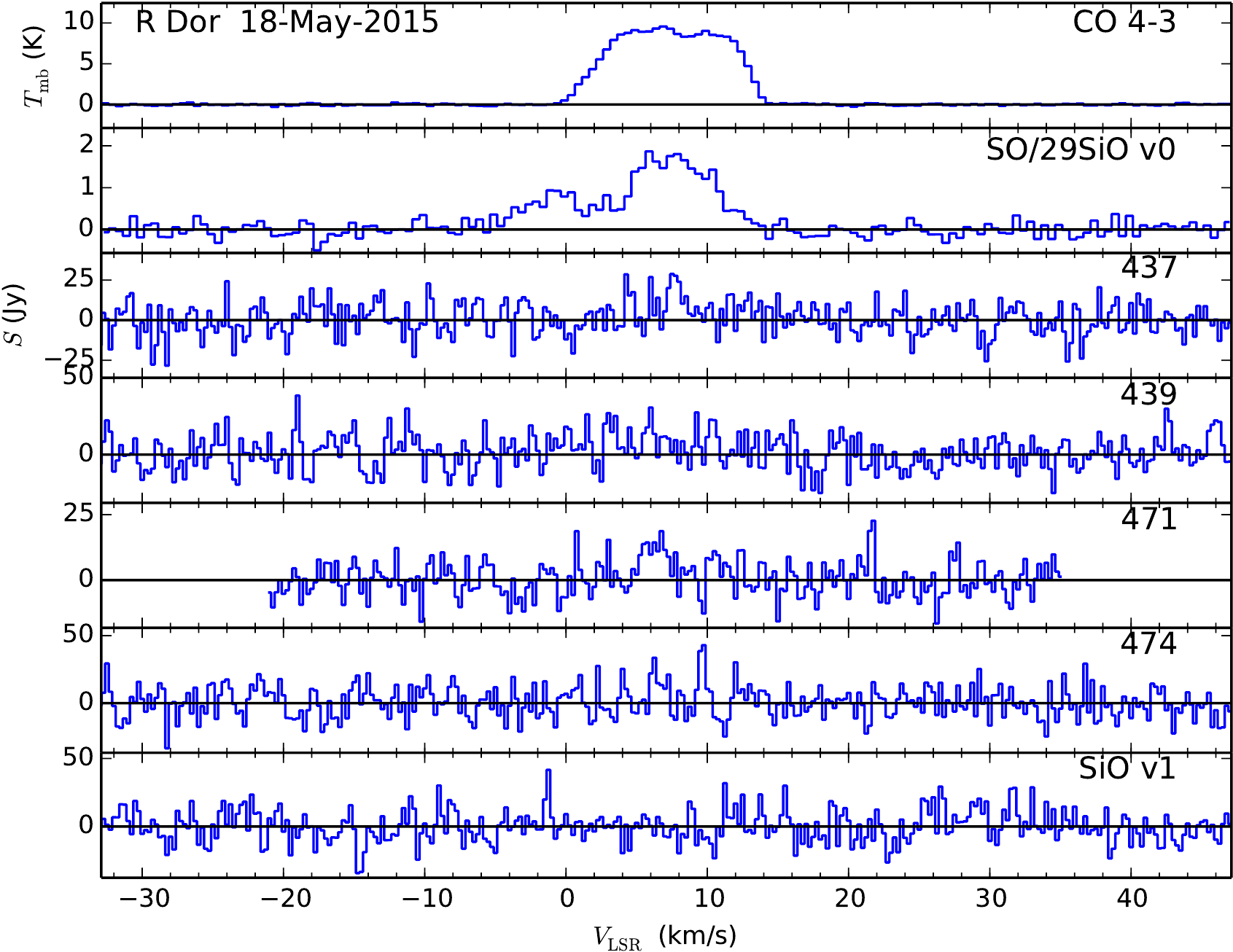} }
     \caption{APEX-3 observations towards R Dor.}
               \label{rdor}       
\end{figure}

\begin{figure*}
   \centering
    \includegraphics[width=17cm]{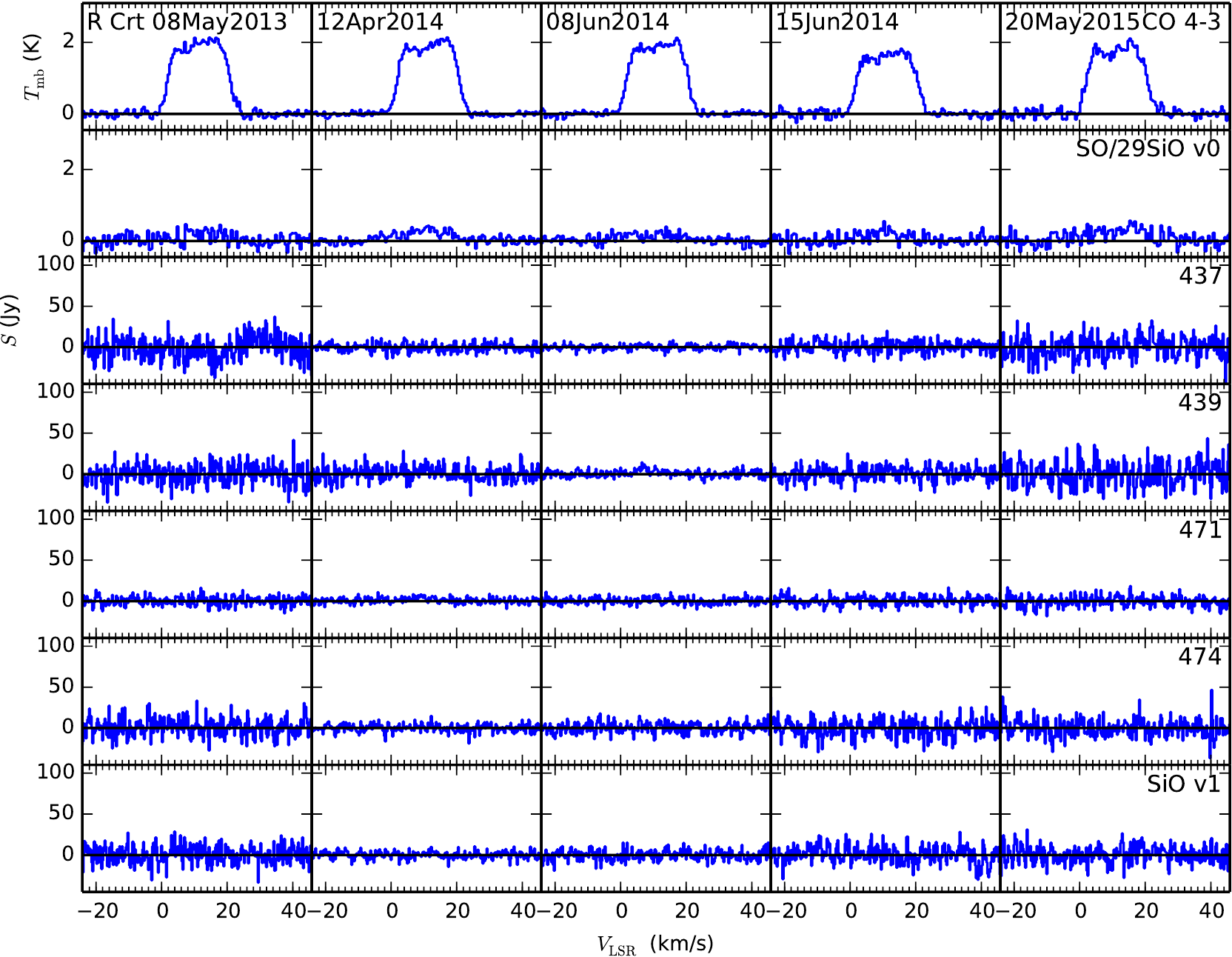}
     \caption{APEX-3 observations towards R Crt.}
               \label{rcrt}       
\end{figure*}

\begin{figure*}
   \centering
    \includegraphics[width=17cm]{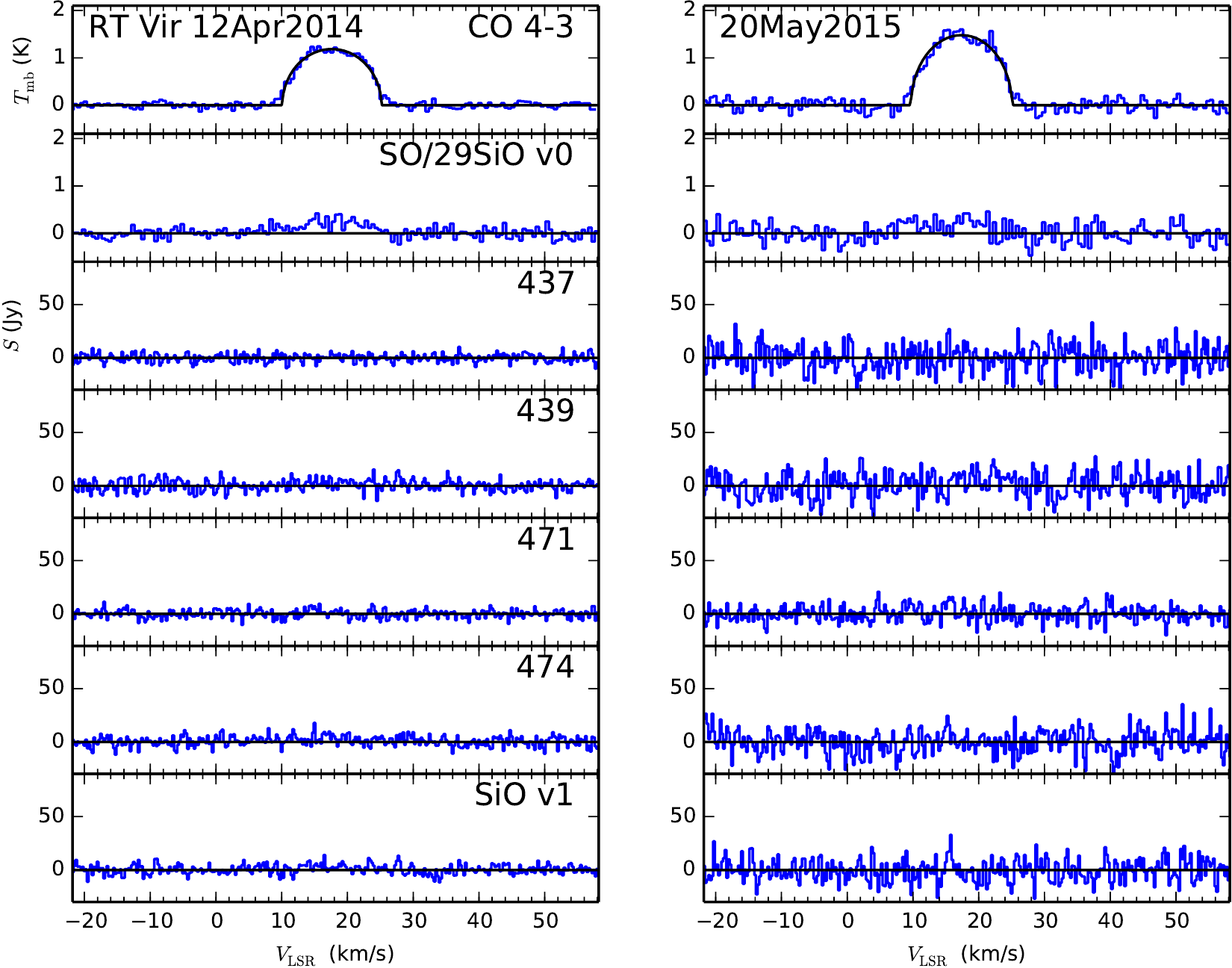}
     \caption{APEX-3 observations towards RT Vir.}
               \label{rtvir}       
\end{figure*}

The CO and water observations are displayed in Figs.~\ref{vxsgr} to \ref{rtvir}.
In addition, blended SO with $^{29}$SiO, and SiO serendipitous detections
are shown. The CO, SO, and $^{29}$SiO observations are reported in main beam temperature scale.
In some cases, a function proportional to $\sqrt{1-(v-v_0)^2/v_e^2}$ has been
fitted to the CO spectra to obtain an estimate of the gas terminal velocity $v_e$.
The water and SiO $v=1$ observations are reported in flux density. The visualised thermal line spectra have
been resampled to a resolution of $0.5\,\mathrm{km\, s^{-1}}$ while the maser lines are shown
with a velocity resolution of $0.25\,\mathrm{km\, s^{-1}}$.
The detections and their characteristics are reported in Tables~\ref{vxsgr_results} to~\ref{rtvir_results}, using a velocity resolution of $0.25\,\mathrm{km\, s^{-1}}$
.

\section{Water maser detections}
\label{sect_water_detect}

In the discussion that follows, we consider maser lines to be detected ('Y' in Table~\ref{sample})
when peak or integrated emission values exceed $3\sigma$, see Tables~\ref{vxsgr_results} to~\ref{rtvir_results}.

There are some caveats to consider when drawing general conclusions from the maser results.
Firstly, the masers are time-variable and for the majority of targets we only
have information at one or two epochs (with the exception of semi-regular variable R Crt, for which
five epochs of data were obtained). In addition, the upper limits in the observations are relatively high, 
and differ between targets and observation epochs.
Nevertheless, with the above cautions we tentatively note the following:

\begin{itemize}

\item {\it VX Sgr and U Her:} These are the highly variable ($\Delta$V $>$ 7.0) longest-period pulsators (P $>$ 400 days).
Strong emission from all four lines is detected. The maser at 437 GHz is significantly stronger than
those at 439, 471, and 474 GHz.

\item {\it RR Aql, W Hya, R Hya, RS Vir, R Leo, and R Aql:}  These are the medium-period pulsators (200 $<$ P $<$ 400 days) with a
magnitude variation of $\Delta$V $>$ 4.0. There are a mixture of detections and non-detections towards
these targets. When detected, the 437 GHz maser is usually the strongest line. However, for W Hya (lowest
visual variation in the group) this is not the case. At the first epoch, detected maser lines have very similar peak
strength and at the second epoch the 474 GHz line is the strongest. W Hya generally seems to be different from the
other targets in that at both epochs it displays $^{29}$SiO $v=0$ $J=11-10$  emission. At the second epoch, it also
displays SiO $v=1$ $J=11-10$ maser emission.

\item {\it R Dor, R Crt, and RT Vir:} These are the semi-regular variables with periods of less than 200 days and a magnitude
variation of $\Delta$V $< 2.0$. The targets in this category appear to be the least likely to host strong water maser emission
at 437, 439, 471, and 474 GHz. R~Crt was observed at five epochs with no detections and RT Vir at two epochs with no detections.
Towards R Dor, there is a marginal detection of the 471 GHz line only.

\end{itemize}

\section{Water maser radiative transfer modelling}
\label{sect:ali}

We performed two different types of radiative transfer modelling.
The first approach was to run a grid of homogeneous models designed to determine the parameter space in
which 437, 439, 471, and 474 GHz maser emission occurs. This approach provides a good indication of
the optimum combination of physical conditions needed to yield strong maser emission in the lines.
The second approach was to model water maser emission from the CSE of a star with a
smooth stellar wind and a range of mass loss rates.
This approach provides a good approximation of the conditions in the outer CSE.
We performed the smooth wind maser modelling to confirm
our interpretation of the grid results and understand the effect of different mass loss rates
on maser emission from the outer CSE. In a future study, we intend to explore this model more thoroughly
by including line overlaps and to treat the inner envelope more accurately with the inclusion of shocks.

The water maser radiative transfer modelling and the CO modelling (see App.~\ref{app:CO}) are based on an
Accelerated Lambda Iteration (ALI) code specifically developed for handling the large positive and negative optical depths encountered
when studying the water emission from molecular clouds and CSEs. The ALI scheme adopted here is
based on the formalism described by \cite{RyHu91}. Also, we include a dust continuum in the radiative transfer following the
approach as outlined in \citet{RyHu92}. Although \citet{RyHu92} also fully describe how to include line overlap in their ALI formalism, we have not enabled this feature in any of our models presented here. This particular ALI code, adapted to spherically symmetric geometry, has been used extensively since the work by \cite{Justtanont2005} and
was tested against other radiative transfer codes by \cite{Maercker2008}. Normally, the iterative procedure starts with a number of lambda iterations (LIs) to provide a good starting point for subsequent accelerated iterations. This procedure was adopted in the homogeneous models but not for the CSE models (see
Sect.~\ref{sect:smooth-wind}) in which accelerated iterations were employed from the first iteration. Other numerical improvements to the convergence and stability of level population changes over iterations can
be invoked if needed. These are Ng-accelerations \citep{Ng1974} or a simple reduction in the amount of level population change between iterations. As demonstrated by \citet{Yates1997} in their ALI plane-parallel code,
one can ensure convergence even in the case of negative optical depths by  sufficiently reducing the step-size. Here, we adopt a slightly different strategy, limiting the acceleration for those transitions that show substantial negative optical depths (typically if the optical depth $< -1$ across a shell).
This is implemented in the code by not including parts of a line where the optical depth is below this limit when integrating
the averaged approximate lambda operator over angles and frequencies.
That is, for those relatively few strongly inverted transitions, we use this modified ALI scheme which would be equal to the standard LI scheme if there
were no contribution at all to the averaged approximate lambda operator.
This step is performed at every iteration when the ALI scheme is employed in order to improve the numerical stability. Using this strategy, the number of radial
points can be lower, but the number of shells should still be large
enough to describe the radial behaviour of the physical conditions in CSEs reasonably well.

The radiative transfer modelling is done separately for the {\it ortho} and {\it para} symmetries of water. For $o$-water, we can
include up to 411 levels and 7597 radiative transitions. The same numbers for $p$-water
are 413 and 7341, respectively. These numbers of levels correspond to an upper energy of about 7200 K and include vibrational levels from both the excited bending mode $\nu_2$ and the symmetric stretching mode $\nu_1$.
For the $\mathrm{H_2}-\mathrm{H_2O}$ collisional data we use
the coefficients of \cite{Faure2008} which are tabulated at 11 kinetic temperatures in the range from 200 K to 5000 K. Linear
interpolation of downward rates is used for temperatures in between the tabulated values. If the
temperature is below 200 K we scale the downward collisional rates at 200 K with $(T/200\,\mathrm{K})^{1/2}$.
All upward collisional rates are subsequently calculated from the downward rates by detailed balance.

\subsection{Physical conditions for 437, 439, 471, and 474 GHz water maser emission}
\label{sect:homogeneous}

We first investigate the physical circumstances in which we obtain maser emission for our 400 GHz target lines. As the homogeneous model cloud, we use a
spherical cloud with a radius of $R= 10^{14}\,\mathrm{cm,}$  a water abundance of $X_\mathrm{H_2O}=5.0\times 10^{-4}$ , and an {\it ortho}-to-{\it para} ratio of three. This high
water abundance is close to the maximum possible large-scale water abundance (as limited by the availability of oxygen atoms). By reducing the water abundance, one can increase
the cloud radius to obtain essentially the same result.
The turbulent velocity is $1\,\mathrm{km\, s^{-1}}$ and no systematic velocity field is used. Although we adopt constant physical conditions throughout the model cloud we still divide the cloud into 15 concentric shells of equal
thickness to allow for radial excitation variations. Optionally, we can include a central source (a black body of a certain luminosity and temperature). The number
of frequency points is 17 and the number of angles (over which the radiation field is integrated) is increased from 12 in the innermost shell to 16 for
the outermost shell. The mean intensity for all transitions at each radial point includes contribution from the line itself emerging from other parts of the cloud,
and thus saturation effects  are always included for inverted transitions. Normally, we include a dust component in each shell, with a dust temperature independent of the gas kinetic temperature and a certain gas-to-dust
mass ratio (typically $R_\mathrm{gd}=100$). For the study presented here, a very simple dust model is adopted with a dust opacity that varies with frequency according to
$\nu^{2}$ and a mass absorption coefficient of $\kappa_{250\mu\rm m}=10\,\mathrm{cm^{2}\, g^{-1}}$. In addition, the model clouds are always exposed to the
2.7 K black body radiation field from the outside.

\begin{figure}
 \resizebox{\hsize}{!}{ \includegraphics[width=12cm]{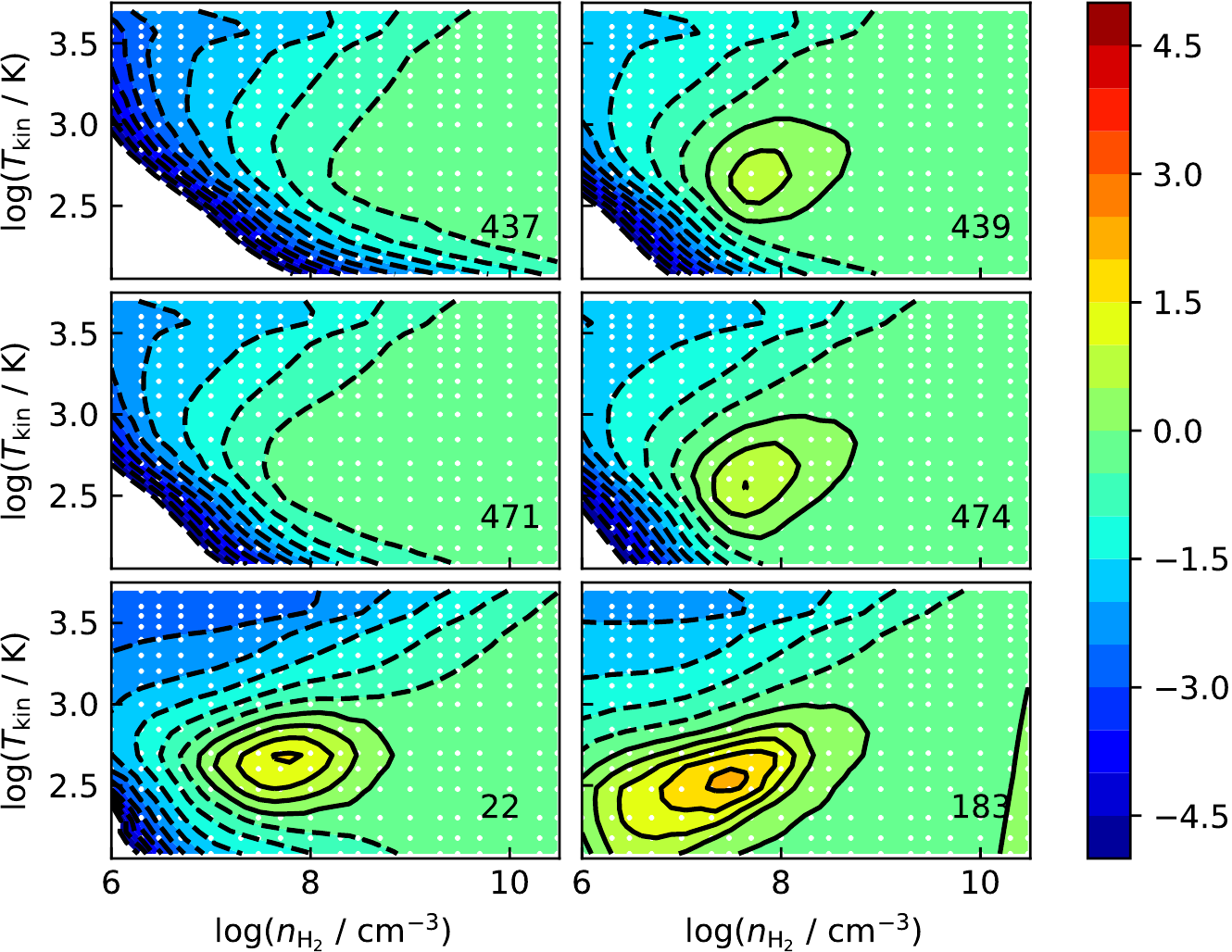} }
  \caption{Homogeneous cloud water grid models with $R_\mathrm{gd}=100$ and $T_\mathrm{dust} = T_\mathrm{kin}$ and no central source. The white
  dots indicate the grid points for which models were run. The transition frequency is shown in each panel in gigahertz. The
  line intensities for each line are described by a base-10 logarithmic contour and colour scale which is normalised to the 557 GHz line intensity ($I_\nu/I_{557}$). The green area depicts the
  region where the line in question has a similar strength to the 557 GHz line. The maximum intensity of the 183 GHz line, which occurs around $T_\mathrm{kin} = 300\,\mathrm{K}$ and
  $n(\mathrm{H_2})=3\times 10^7\,\mathrm{cm^{-3}}$, is about $10^2$ times stronger than the 557 GHz line intensity. }
            \label{homogeneous_std}
  \end{figure}

The ALI code is run over a grid of values. We vary the kinetic temperature, $T_\mathrm{kin}$, from 120 K to 5000 K and the H$_2$ density, $n(\mathrm{H_2})$, is varied from $10^6\,\mathrm{cm^{-3}}$ to
$3\times 10^{10}\,\mathrm{cm^{-3}}$.
For each set of parameters, we compute a line profile for several different transitions, always adopting a beam size equal to the cloud diameter.
Our main aim here is to investigate how strong the 437, 439, 471, and 474 GHz maser emissions can become. To visualise this, we normalise the integrated intensity,
$I_\nu$, of each
line to that of the 557 GHz {\it ortho} line (i.e. we compute the base-10 logarithm of $I_\nu/I_{557}$). The results for the case where $R_\mathrm{gd}=100$ and $T_\mathrm{dust} = T_\mathrm{kin}$ without a central source are shown in
Fig.~\ref{homogeneous_std}. Besides our 400 GHz lines we also include the results for the 22 GHz and 183 GHz water lines for comparison. As can be seen, the
439 and 474 GHz lines are  weakly masing while the 437 and 471 GHz lines show no masers.

\begin{figure}
 \resizebox{\hsize}{!}{ \includegraphics[width=12cm]{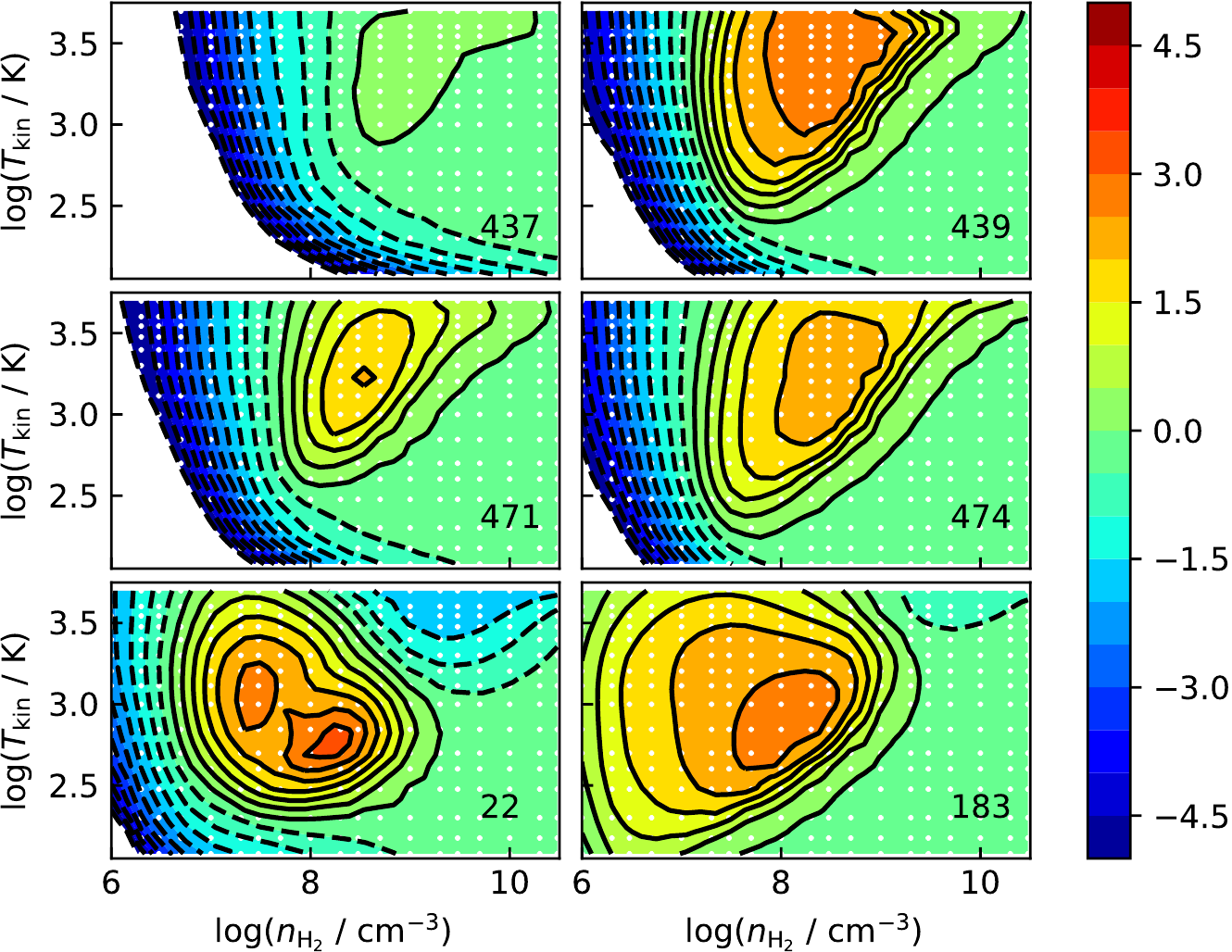} }
  \caption{Homogeneous cloud water models without a dust component and no central source. Scales and contours are as in Fig.~\ref{homogeneous_std}. }
            \label{no_dust_std}
  \end{figure}

The inclusion of dust has a strong impact on the maser activity \citep{Yates1997}. In Fig.~\ref{no_dust_std}, we show the previous model set but this time not including any dust
(or central source). If we compare with the results in Fig.~\ref{homogeneous_std}, we see that the inclusion of dust (with $T_\mathrm{dust} = T_\mathrm{kin}$) has a clear quenching effect on the maser action.

\begin{figure*}
\centering
  \includegraphics[width=17cm]{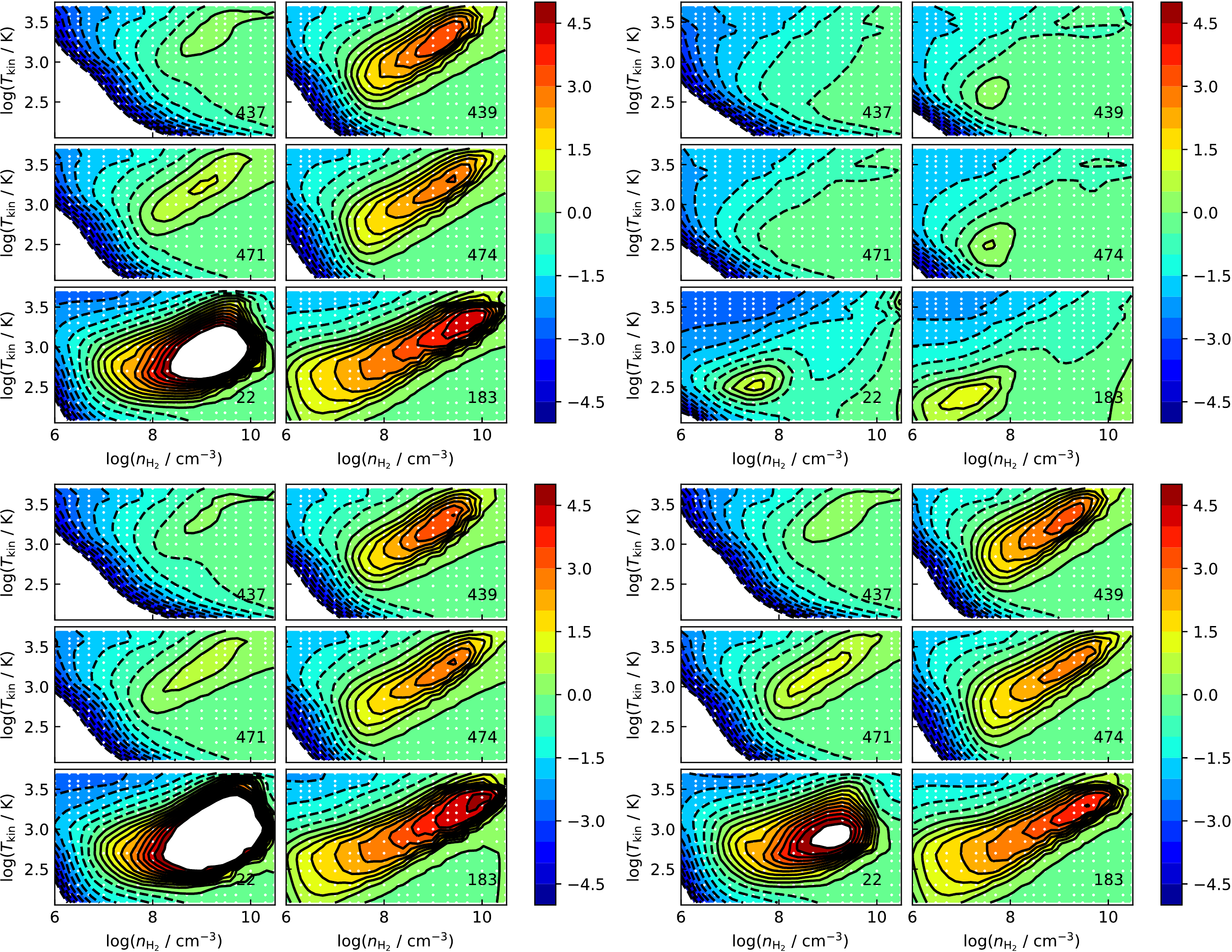}
  \caption{Homogeneous cloud water models for different dust properties.
  Top left: $R_\mathrm{gd}=100$ and $T_\mathrm{dust} = T_\mathrm{kin}/2$.
  Top right: $R_\mathrm{gd}=100$ and $T_\mathrm{dust} = 3T_\mathrm{kin}/2$.
  Bottom left: $R_\mathrm{gd}=50$ and $T_\mathrm{dust} = T_\mathrm{kin}/2$.
  Bottom right: $R_\mathrm{gd}=200$ and $T_\mathrm{dust} = T_\mathrm{kin}/2$.
   Scales and contours are as in Fig.~\ref{homogeneous_std}. In the 22 GHz maser case, intensity ratios, $I_{22}/I_{557}$, are sometimes in excess of $10^5$ (central white areas).
   The strongest 22 GHz maser in these four cases occurs in the bottom left panel for $T_\mathrm{kin} = 700\,\mathrm{K}$ and
   $n(\mathrm{H_2})=10^{10}\,\mathrm{cm^{-3}}$. Here the integrated intensity ratio is about $10^7$.
}
         \label{four_models}
\end{figure*}

Following the results of \citet{Yates1997}, in that some masers become stronger when $T_\mathrm{dust} < T_\mathrm{kin}$, we ran the models in Fig.~\ref{homogeneous_std}
but this time with $T_\mathrm{dust} = T_\mathrm{kin}/2$ and $3 T_\mathrm{kin}/2$. We also ran models with $R_\mathrm{gd}=50$ and 200 for the case where
$T_\mathrm{dust} = T_\mathrm{kin}/2$. These four model runs are displayed in Fig.~\ref{four_models}. Allowing for $T_\mathrm{dust} = T_\mathrm{kin}/2$ leads to
a remarkable increase of the maser line intensities by several orders of magnitude. Increasing the amount of dust with $R_\mathrm{gd}=50$ only leads to a
moderate increase in the 400~GHz maser line intensity. The 22~GHz maser intensity behaves differently and is strongly dependent on
the amount of dust. This is also valid for the 183~GHz maser intensity but to a lesser extent.

The key factor here for strong masers seems to be that the dust temperature is lower than the kinetic temperature.
In fact, in the case of our model, setting $T_\mathrm{dust} \sim T_\mathrm{kin}/3$ maximises the maser action. On the other hand, for $T_\mathrm{dust} \ga T_\mathrm{kin}$ the
maser action is rather quenched. In our homogeneous models we have not seen any marked deviations from $I_{437} \la I_{471} \la I_{474} \la I_{439}$. For
the latter pair of lines, we conclude that although the maximum peak temperature is normally somewhat higher for the 439~GHz line than for the 474~GHz line, the density--temperature
region over which masers occur is slightly different in the 474~GHz case than in the 439~GHz case (see Fig.~\ref{four_models}). As a result, line ratios of these
maser transitions may behave differently depending on the exact nature of any gradients in the physical parameters.

We also tested the effect of the inclusion of a central black body with a radius of $r=0.1R$ and black body temperatures up to 4000 K, but only see minor changes to the intensities of the 400 GHz masers.

\subsection{Circumstellar smooth-wind maser modelling}
\label{sect:smooth-wind}

To model the water emission from CSEs we divided our model cloud with radius $R$ into a larger number of shells (typically 60
with logarithmic spacing). Both for {\it ortho} and {\it para} species, only levels below 4000~K are included. A few test runs with
a larger number of shells (120) and higher energy limit (6000~K) resulted in minor changes: less than 5\% variation in maser peak
intensities for the targeted 400 GHz lines.
The innermost radius, $R_\mathrm{i}$, is given by the luminosity and temperature of the star
(see Table~\ref{sample}). The kinetic temperature is always set to 33 K at the cloud surface (as the water radius
is only 0.6 of the CO radius where $T_\mathrm{kin}=20\,\mathrm{K}$) and to the stellar
temperature at $R_\mathrm{i}$. The kinetic temperature decline in the shells then follows $r^{-\alpha}$ with
$\alpha$ near 1 (0.68 for U~Her and 0.83 for W~Hya). The H$_2$ density radial dependence is entirely given here via the continuity equation; by an adopted
mass loss rate, ${\dot M}$, and a radial gas expansion velocity according to
\begin{equation}
   v_\mathrm{e}(r) = v_\mathrm{i} + \left(v_R - v_\mathrm{i}\right) \left(1 - \frac{R_\mathrm{i}}{r}\right)^\beta,
\end{equation}
where $v_\mathrm{i}$ is the radial gas velocity at $R_\mathrm{i}$ and $v_R$ is the terminal gas velocity. Following
\citet{Maercker2016}, here we  set $v_\mathrm{i}= 3\,\mathrm{km\,s^{-1}}$ and use $\beta=3/2$. As in the homogeneous models, the turbulent
velocity width is $1\,\mathrm{km\,s^{-1}}$.

The water abundance due to photo-dissociation in the interstellar radiation field \citep[cf.][]{Netzer1987, Maercker2016}
can be described by
\begin{equation}
   X_\mathrm{H_2O}(r) = X(0) \exp\left(-(r/r_e)^2\right),
\end{equation}
where $X(0)$ is the water abundance near the centre and $r_e$ is the $e$-folding radius describing the relatively sharp decline
of water molecules due to photo-dissociation in the outer part of the CSE. Here, as in the previous section, we set $X(0)$ to
$5\times 10^{-4}$ and $r_e = 0.8 R$. Also, we use  CSE CO radius $R=R_{1/2}$ (see App.~\ref{app:CO} and Table~\ref{co_modelling}) as the outer water radius. For comparison we also
include the modelled profiles for $r_e = 0.1 R$. This latter $r_e$ is more
in line with the water photo-dissociation radius as formulated by \citet{Maercker2016} and based on the work by \citet{Netzer1987}.
We use a relatively large value of $X(0)$   here because we want to explore conditions in which water maser activity is most prone to occur.
In our water modelling, we mainly vary the mass loss rate from $3\times 10^{-8}$ to $10^{-4}\,\mathrm{M_\sun\, yr^{-1}}$ and
we adopt the same dust properties as in the previous section, except that we use $R_\mathrm{gd}=200$ and $T_\mathrm{dust}(r) = T_\mathrm{kin}(r)/2$.

Two model cases were run. The sources selected were U~Her and W~Hya because they both show 400 GHz masers while having quite
different sizes for their CSEs; see Table~\ref{co_modelling}.

\begin{figure}
\resizebox{\hsize}{!}{ \includegraphics[]{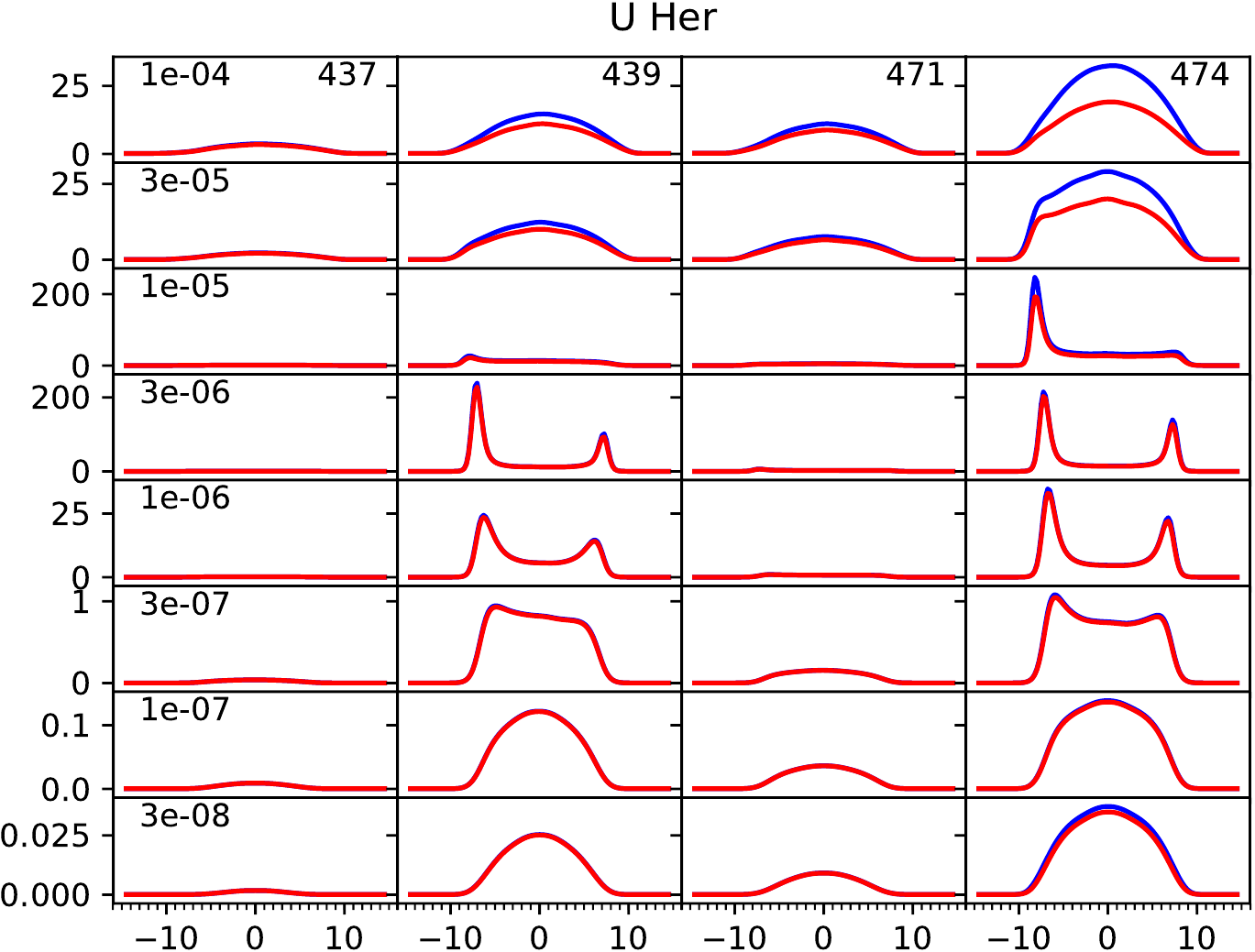} }
\caption{Model water spectra of the 400 GHz lines in the case of U Her. The frequency is shown in the top row spectra in gigahertz. Each row of spectra
  corresponds to the mass loss rate indicated
  in the leftmost spectrum. The blue profiles correspond to $r_e = 0.8 R$ and the red profiles to $r_e = 0.1 R$. The
  vertical scale, in Janskys, is different for each mass loss rate. The horizontal scale is in velocity ($\mathrm{km\, s^{-1}}$). The strongest maser activity is found for
  $\dot{M} = 3\times 10^{-6}\,\mathrm{M_\sun\, yr^{-1}}$ and is absent for the highest mass loss rates. The mass loss rate derived
  from the CO lines is $2\times 10^{-7}\,\mathrm{M_\sun\, yr^{-1}}$.}
\label{uher_400GHz_models}
\end{figure}

The U~Her model results, $R={1.3\times 10^{16}\,\mathrm{cm}}$ (see Table~\ref{co_modelling}), are shown in Fig.~\ref{uher_400GHz_models} for
different mass loss rates. The model cloud spectra haven been convolved with Gaussian beams appropriate for the APEX telescope.
The masers (for 439 and 474 GHz) appear only for mass loss rates around $(1-10)\times 10^{-6}\,\mathrm{M_\sun\, yr^{-1}}$.
Here the blueshifted peak is stronger than the redshifted peak since much of the redshifted amplification path at $\sim +v_e$
is obscured by the central opaque source and the stellar radiation is amplified for $\sim -v_e$ \citep{Gonzalez1999}.
The radius range, in which the radial optical depths in the 439 GHz line are negative, is [$0.006 R$, $0.081 R$] for a mass loss
of $3\times 10^{-6}\,\mathrm{M_\sun\, yr^{-1}}$ at which the 439 GHz line is the strongest line (having a peak intensity around
250~Jy at the blueshifted peak). For other mass loss rates the 474 GHz transition is the dominant line.

The mass loss rate for U~Her determined by the CO modelling is $2.0\times 10^{-7}\,\mathrm{M_\sun\, yr^{-1}}$; see
Table~\ref{co_modelling}. No strong 400 GHz masers at $\pm v_e$ are expected for this mass loss rate, even with $r_e = 0.8 R$.
The expected 474 GHz peak intensity for the non-masing line at $\dot{M} = 3\times 10^{-7}\,\mathrm{M_\sun\, yr^{-1}}$ is
around 1~Jy which is below the observed rms level for U~Her (see Table~\ref{uher_results}).

\begin{figure}
 \resizebox{\hsize}{!}{  \includegraphics[]{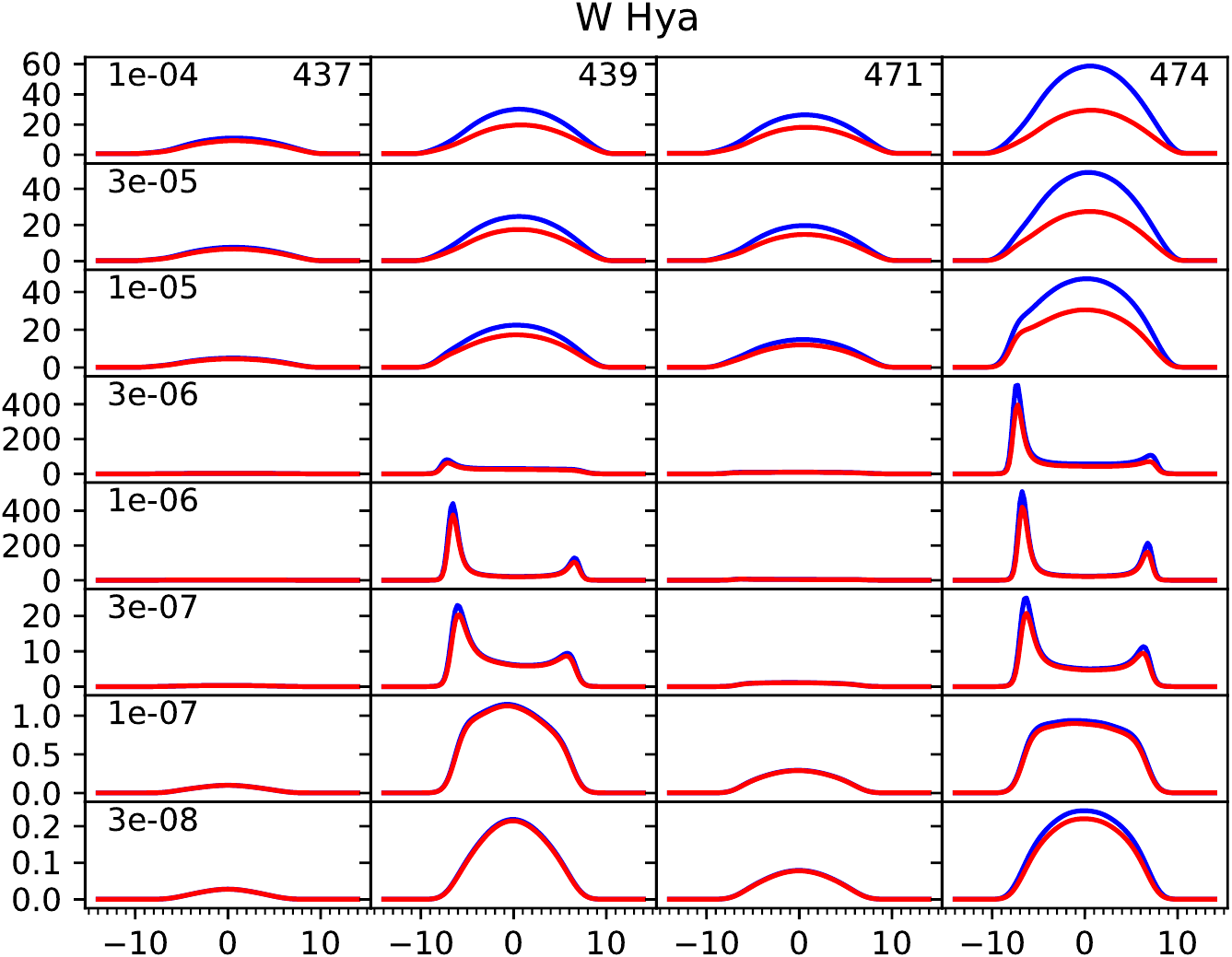} }
\caption{Model water spectra of the 400 GHz lines for W Hya. The layout and scales are as in Fig.~\ref{uher_400GHz_models}. The strongest maser activity is found for $\dot{M} = (1-3)\times 10^{-6}\,\mathrm{M_\sun\, yr^{-1}}$. The mass loss rate derived from the CO lines is $5\times 10^{-7}\,\mathrm{M_\sun\, yr^{-1}}$.}
\label{whya_400GHz_models}
\end{figure}

Figure~\ref{whya_400GHz_models} shows the W~Hya model 400 GHz spectra for the same set of mass loss rates as for U~Her.
The W~Hya CSE, at ${5\times 10^{15}\,\mathrm{cm}}$, is smaller than that of U~Her but its mass loss rate, at $3.2\times 10^{-7}\,\mathrm{M_\sun\, yr^{-1}}$,
is similar to that of U~Her. The masers show the same behaviour for W~Hya as in the case of our U~Her models. For a mass loss rate of $1\times 10^{-6}\,\mathrm{M_\sun\, yr^{-1}}$ , the
439~GHz maser activity is maximised and the radial range over which the radial optical
depths are negative is [$0.006 R$, $0.11 R$].

For our homogeneous models in Sect.~\ref{sect:homogeneous}, we find that the peak brightness of the 439~GHz line is usually stronger
than that of the 474~GHz line. On the other hand, in the case of the U~Her and W~Hya models, we see that the 439~GHz line
is as bright as the 474 GHz line, albeit only at a specific mass loss rate ($3\times 10^{-6}\,\mathrm{M_\sun\, yr^{-1}}$ for U~Her and
$1\times 10^{-6}\,\mathrm{M_\sun\, yr^{-1}}$ for W~Hya). Nevertheless, the 474~GHz maser activity extends over a larger range of mass loss rates.
From the results in Fig.~\ref{four_models} we see that the optimum combination for 439 GHz masers to occur is given
by $T_\mathrm{kin} \sim 10^{0.3 \log(n(\mathrm{H_2}) / \mathrm{cm^{-3}}) + 0.5}$. If we adopt this kinetic temperature behaviour
at each radius of our model cloud we obtain much stronger masers (in all 400 GHz lines). Likewise, if we start with our kinetic temperature relation
$T_0 (r/R)^\alpha$ and use the optimum maser relation to determine the H$_2$ density, we obtain essentially the same result.

 \section{Discussion}
 
 Out of our sample of 11 evolved stars, 7 display one or more of the maser lines
 studied here at 437, 439, 471, and 474 GHz. We therefore find that these masers are common
 in CSEs of evolved stars. Moreover, we note our study is only sensitive to masers
 of peak flux density $\ga$15 Jy, and so they may be even more common at for example the 1 Jy level. As noted earlier in
 Sect.~\ref{sect_water_detect}, we tentatively find that the presence of strong masers in these lines
 correlates with the variability in the magnitude of the star during the pulsation cycle. This could
 indicate that shocks are needed to create the pumping conditions for the masers. It is
 plausible that one must provide a better physical description of the inner part of the CSE, where dust particles form and
 help accelerate the gas, to understand the appearance of 400 GHz water masers.
 This is most likely coupled to the pulsation of the central star,
 and models like those investigated by \citet{Bladh2019} could provide further insight in this area.
 
 One of the more striking observational results is that the 437 GHz maser line is typically by far the strongest.
 However, this transition is not found to be
 significantly inverted in the work of \citet{Gray2016},  which predicts many other of the
 common masers. It is also not found to be strongly masing in the current work, despite the
 fact that the grid of physical conditions covers all those expected for the likely location of the
 masers (the masers are unlikely to predominantly originate in the outer wind region because the
 line shapes would then be double-peaked and bracket the stellar velocity). We therefore speculate
 that some other ingredient must be necessary for the production of strong maser emission
 in this line. We suggest line overlap may be the cause and therefore we plan to perform a systematic investigation
 of this in future work. \citet{Gray2016} did include line overlap within the of {\it para} and {\it ortho}
 water species separately, but it may be that overlap between the species is 
also of importance
 here. One such overlap involves the rotational transitions $J_{K_a,K_c} = {6}_{6,1}-{5}_{5,0}$ ({\it o}) and ${6}_{6,0}-{5}_{5,1}$ ({\it p}) which
 are within $3.6\,{\mathrm{km\, s^{-1}}}$ of each other around $33.0\,\mathrm{\mu m}$. Interestingly, this particular line overlap is directly coupled to three of our four 400 GHz maser transitions (see Table~\ref{linelist}). Another {\it para} and {\it ortho}
 overlap involving the lower 437~GHz level is the $J_{K_a,K_c} = {7}_{7,1}-{6}_{6,0}$ ({\it p}) and ${7}_{7,0}-{6}_{6,1}$ ({\it o}) transitions
 at $28.6\,\mathrm{\mu m}$. This line overlap corresponds to $0.4\,\mathrm{km\, s^{-1}}$.
 We also note that there is an overlap at $6.63\,\mathrm{\mu m}$ of three ro-vibrational transitions within $10.4\,\mathrm{km\, s^{-1}}$ that  are
 related to the 437 GHz transition. These are the $\nu_2=1$ to 0 transitions ${6}_{5,2}-{6}_{6,1}$ ({\it o}), ${6}_{5,1}-{6}_{6,0}$ ({\it p})
, and ${7}_{5,3}-{7}_{6,2}$ ({\it p}). We used frequencies listed in the
 high-resolution transmission molecular absorption database \citep[HITRAN,][]{Gordon2017} to calculate the line overlap distances.
 Since the {\it ortho}-species is likely to be more abundant than the {\it para}-species (3:1 at thermal equilibrium), the line
 overlap between the species could generally be more important for {\it para} lines. However, for individual lines, the exact details of
 the overlap, such as for example the amount of overlap (including velocity field) and $A$-coefficients of involved transitions, will be paramount.
 We note that the velocities of up to $>$ 10 km s$^{-1}$  required for the overlaps may explain why
 the 437 GHz maser has not yet been found to be associated with star-forming regions. In any event, a more detailed investigation of possible line overlaps for water may be of
 importance to better understand the mechanism(s) giving rise to 437~GHz masers.
 
 An interesting feature of these masers is the significance of dust, which was previously indicated
 by the work of \citet{Yates1997}.
 Figures~\ref{homogeneous_std} and~\ref{no_dust_std} indicate that the maser lines in general become stronger when
 there is no dust, compared with when there is dust present at the same temperature
 as that of the gas and the maser action is quenched. Figure~\ref{four_models} displays the line intensities for four
 different dust temperatures, showing that maser emission is strong when the
 dust temperature is lower than that of the gas temperature. The optimum 
 situation for strong maser emission is when the dust temperature is about one-third of
 the gas temperature for our homogeneous cloud models. The amount of dust also affects
maser activity for the 400 GHz masers, although the impact is smaller than that on
 the 22~GHz maser; see Fig.~\ref{four_models}. Lowering the gas-to-dust ratio from
 200 to 50 seems to slightly increase the maser activity for the 439~GHz and
 474~GHz maser lines while 437~GHz and 471~GHz maser activity goes in the
 opposite direction and is slightly diminished.
 
 For water masers at 22 and 183 GHz \citep{Gonzalez1998}, it is known that in objects
 of relatively low mass loss rate, the maser emission is tangentially beamed from the
 inner CSE where radial velocity gradients are too high to meet the
 so-called velocity coherence required for maser amplification. For objects of higher
 mass loss rate, the physical conditions needed to pump the masers can be found further
 from the star where the radial velocity gradients are lower. Towards these objects,
 the maser line shapes are no longer predominantly single peaked at the stellar velocity,
 but have a double-peaked structure bracketing the stellar velocity at up to the terminal
 velocity for the object. In some cases, the redshifted emission is obscured by the star,
 so only the blueshifted peak is observed at a velocity offset from the stellar velocity.
 
 For our sample, all of the observed 400 GHz lines appear to be originating predominantly from the inner
 CSE (the acceleration zone region), that is, they occur near the velocity centroid of the CO emission. This
 is quite the opposite to what is indicated by the smooth wind modelling.
 The smooth wind modelling indicates that emission can, as at 22 and 183 GHz and perhaps 
 for all water masers, originate from the outer wind; see for example \citet{Gonzalez1999}. From our
 U~Her and W~Hya modelling (assuming the preferable conditions for maser emission: high water abundance
 of $5\times 10^{-4}$ and $T_\mathrm{dust}(r) = T_\mathrm{kin}(r)/2$), a strong and double-peaked maser profile is achieved for
 the 437 and 474 GHz lines when the mass loss rate is
 in the range of about $10^{-6} - 10^{-5} \,\mathrm{M_\sun\, yr^{-1}}$. For a higher mass loss rate, the maser action becomes quenched.
 With the exception of VX~Sgr and possibly R~Aql, none of our targets has such a high mass loss rate.

\section{Summary and conclusions}
 
We performed observations at the  water maser frequencies of 437, 439, 471, and 474 GHz using
APEX towards 11 stars. Our findings can be summarised as follows.

\begin{itemize}

\item The masers are common and strong towards Mira variables. We suggest that the masers
are either less strong or are highly variable and absent towards semi-regular variables for some part of the
stellar pulsation cycle.

\item The line shapes of the observed maser lines are approximately centred at the stellar velocity rather than
being double-peaked and bracketing the stellar expansion velocity. This likely indicates
that the masers originate predominantly from the the wind acceleration
zone of the CSEs rather than from the outer wind.

\item The 437 GHz line is typically by far the strongest line in the observations. We cannot
reproduce this behaviour with radiative transfer models. In our homogenous models, 
there is never significant deviation from $I_{437} \la I_{471} \la I_{474} \la I_{439}$. In our CSE models,
the maser activity of the 474~GHz line occurs over a larger range of mass loss rates than
the 439~GHz maser activity. Also, no strong maser activity in the 437 and 471 GHz transitions
is seen in our CSE models.

\item We speculate that line overlap both within and
between the {\it ortho} and {\it para} water species may need to be incorporated to reproduce
observations of the 437 GHz line. Purely rotational overlaps around 30 $\mu$m and ro-vibrational overlaps at 6.63 $\mu$m may be
the key.

\item We find that dust temperature plays an important role in the production of
strong masers in these lines. The optimum dust temperature for strong maser
emission is around $T_\mathrm{kin}/3$ for homogeneous models. It is clear that the gas and dust
interplay is important for these masers and therefore detailed characteristics of the dust properties and the heating and cooling processes of the gas may be important to gain a better understanding of the 400 GHz water masers.

\item Our smooth wind CSE modelling confirms that it is possible to achieve
double-peaked water maser emission for the 439 and 474 GHz lines as is the case for the 22 and 183 GHz water masers. However, it requires  a high water
abundance in the region where the expansion velocity is near the terminal velocity; a mass loss rate in the range of about
$10^{-6} - 10^{-5}\,\mathrm{M_\sun\, yr^{-1}}$ seems to be needed for this to happen. We cannot reproduce the observed
line shapes in the smooth wind modelling, namely single-peaked maser lines at the stellar velocity. We suggest that this
is another indicator that the masers originate from the inner envelopes of our targets and require the presence of
strong shocks. 

\end{itemize}

\begin{acknowledgements}
We thank the OSO observing teams, aided by the APEX staff, for performing these observations,
which were part of the APEX Swedish queue. We are also very grateful to Hans Olofsson for discussions and comments on
an earlier version of this paper.
We acknowledge with thanks the variable star observations from the {\em AAVSO International Database} contributed by observers worldwide and used in this research.
\end{acknowledgements}


\bibliographystyle{aa} 
\bibliography{water_37774} 

\begin{appendix}

\section{Observational results}

In this Appendix, we present the observational measurements towards each target.
The measurements and stellar phases at the different epochs have been entered in
Tables~\ref{vxsgr_results}-\ref{rtvir_results}. If the 437 GHz line is the strongest
line in any source and epoch, its tabulated intensity value is set using a boldfaced font.
All peak and channel RMS values relate to a velocity resolution of $0.25\,\mathrm{km\, s^{-1}}$.
     
\begin{table}
   \caption{\label{vxsgr_results}  VX Sgr results}
   \centering
   \begin{tabular}{lrr}
   \hline\hline 
   &  \multicolumn{2}{c}{May 18 2015 (stellar phase $\sim$ 0.05)}\\
 \multicolumn{1}{c}{Line}               &   \multicolumn{1}{c}{Peak (rms)}      &  \multicolumn{1}{c}{Integrated (rms)} \\
 \hline
 CO 4-3                             & 3.77 (0.36) K                                     & 99.7 (1.3) K kms$^{-1}$\\
 $^{29}$SiO 11-10 v=0     &  ... (0.54) K                                          & \multicolumn{1}{c}{...}\\
 H$_2$O 437 GHz          &  {\bf 354.4 (12.3)} Jy                                    & 1022.0 (45.0) Jy kms$^{-1}$\\
 H$_2$O 439 GHz         &  82.8 (12.5) Jy                                      & 507.8 (46.1) Jy kms$^{-1}$\\
 H$_2$O 471 GHz         & 85.2 (16.4) Jy                                        & 576.0 (60.1) Jy kms$^{-1}$\\
 H$_2$O 474 GHz         & 84.1 (27.1) Jy                                       & 271.9 (99.5) Jy kms$^{-1}$\\
 SiO 11-10 v=1               &  ...    (27.9) Jy                                       &       \multicolumn{1}{c}{...}                 \\            
\hline  
   \end{tabular}
\end{table}

\begin{table}
   \caption{\label{uher_results}  U Her results}
   \centering
   \begin{tabular}{lrr}
   \hline\hline 
  & \multicolumn{2}{c}{April 12 2014 (stellar phase $\sim$ 0.05)} \\
 \multicolumn{1}{c}{Line}               &   \multicolumn{1}{c}{Peak (rms)}      &  \multicolumn{1}{c}{Integrated (rms)} \\
 \hline
 CO 4-3                        &  0.72 (0.073) K     & 8.0 (0.2) K kms$^{-1}$           \\
 $^{29}$SiO 11-10 v=0  &  ... (0.129)  K       & ... (0.4) K kms$^{-1}$             \\
H$_2$O 437 GHz         & {\bf  929.8 (4.0)} Jy     & 1649.0  (11.0)  Jy kms$^{-1}$  \\
H$_2$O 439 GHz         &  38.3 (5.9) Jy       & 218.4 (16.3) Jy kms$^{-1}$     \\   
H$_2$O 471 GHz         & 64.4 (4.3) Jy        & 144.5 (12.0) Jy kms$^{-1}$     \\  
H$_2$O 474 GHz         &  29.9 (5.0) Jy       & 66.7 (13.6) Jy kms$^{-1}$       \\  
 SiO 11-10 v=1              &   ... (5.2) Jy          & \multicolumn{1}{c}{...}              \\
\hline
& \multicolumn{2}{c}{May 20 2015 (stellar phase $\sim$ 0.2)} \\
 \multicolumn{1}{c}{Line}               &  \multicolumn{1}{c}{Peak (rms)}      &  \multicolumn{1}{c}{Integrated (rms)} \\
 \hline
CO 4-3                        &  0.86 (0.101) K                              & 8.9 (0.3) K kms$^{-1}$   \\
 $^{29}$SiO 11-10 v=0  &  ... (0.159)  K                                &  ... (0.5) K    kms$^{-1}$   \\       
H$_2$O 437 GHz         &  {\bf 896.8 (11.2)} Jy                             & 3252.0  (31.0) Jy  kms$^{-1}$   \\
H$_2$O 439 GHz         &  58.8 (4.3) Jy                                   & 148.0  (31.1) Jy kms$^{-1}$   \\   
H$_2$O 471 GHz         & 19.9 (5.5) Jy                                   & 41.6 (15.2) Jy kms$^{-1}$   \\  
H$_2$O 474 GHz         &  24.9 (8.3) Jy                                   & 26.2 (22.8) Jy  kms$^{-1}$   \\  
 SiO 11-10 v=1              &   ... (8.0) Jy                                    &  \multicolumn{1}{c}{...}  \\
 \hline  
   \end{tabular}
\end{table}

\begin{table}
   \caption{\label{rraql_results}  RR Aql results }
   \centering
   \begin{tabular}{lrr}
   \hline\hline 
     & \multicolumn{2}{c}{May 18 2015 (stellar phase $\sim$ 0.8)}\\
 \multicolumn{1}{c}{Line}               &   \multicolumn{1}{c}{Peak (rms)}      &  \multicolumn{1}{c}{Integrated (rms)} \\
 \hline
 CO 4-3                             & 1.19 (0.14) K        & 11.7 (0.4) K kms$^{-1}$ \\
 $^{29}$SiO 11-10 v=0     &  ... (0.20) K           &  \multicolumn{1}{c}{...}\\
 H$_2$O 437 GHz          & {\bf 73.1 (10.5)} Jy          & 64.1 (26.9) Jy kms$^{-1}$\\
 H$_2$O 439 GHz         &  ... (11.6) Jy      &  \multicolumn{1}{c}{...}\\
 H$_2$O 471 GHz         & ... (5.6) Jy  &  \multicolumn{1}{c}{...}\\
 H$_2$O 474 GHz         & ... (8.9) Jy  &  \multicolumn{1}{c}{...}\\
 SiO 11-10 v=1               &   ... (8.3) Jy &  \multicolumn{1}{c}{...}\\
\hline  
   \end{tabular}
\end{table}

\begin{table}
   \caption{\label{whya_results}  W Hya results}
   \centering
   \begin{tabular}{lrr}
   \hline\hline 
  & \multicolumn{2}{c}{April 13 2014 (stellar phase $\sim$ 0.70)} \\
 \multicolumn{1}{c}{Line}               &   \multicolumn{1}{c}{Peak (rms)}      &  \multicolumn{1}{c}{Integrated (rms)}  \\
 \hline
 CO 4-3                        &  4.07 (0.076) K                                                  & 52.7 (0.2) K kms$^{-1}$                   \\
 $^{29}$SiO 11-10 v=0  &   1.34 (0.130) K                                                &   9.3 (0.4)   K kms$^{-1}$                  \\
H$_2$O 437 GHz         &    {\bf 22.1 (4.3)} Jy                                            & 77.3 (11.7) Jy kms$^{-1}$               \\
H$_2$O 439 GHz         &   ... (5.9) Jy                                                         & \multicolumn{1}{c}{...}                      \\
H$_2$O 471 GHz         &  21.2 (3.8) Jy                                                      & 126.8 (10.5) Jy kms$^{-1}$             \\
H$_2$O 474 GHz         &  21.9 (5.7) Jy                                                      & 104.8 (15.8) Jy kms$^{-1}$               \\
 SiO 11-10 v=1              &   ... (5.5) Jy                                                         & \multicolumn{1}{c}{...}                       \\
 \hline  
 & \multicolumn{2}{c}{May 20 2015 (stellar phase $\sim$ 0.75)} \\
 \multicolumn{1}{c}{Line}               &   \multicolumn{1}{c}{Peak (rms)}      &  \multicolumn{1}{c}{Integrated (rms)} \\
 \hline
 CO 4-3                        & 3.97 (0.110) K                                   & 49.6 (0.3) K kms$^{-1}$  \\
 $^{29}$SiO 11-10 v=0  & 1.18 (0.165) K                                  & 9.0 (0.5) K kms$^{-1}$  \\
H$_2$O 437 GHz         & 57.9 (7.1) Jy                                    & 278.5 (19.5) Jy kms$^{-1}$  \\
H$_2$O 439 GHz         & 52.9 (7.5) Jy                                     & 275.2 (20.5) Jy kms$^{-1}$  \\
H$_2$O 471 GHz         & 41.5 (5.3) Jy                                     & 191.4 (14.6) Jy kms$^{-1}$  \\
H$_2$O 474 GHz         & 113.3 (7.4) Jy                                   & 323.9 (20.4)  Jy kms$^{-1}$  \\
 SiO 11-10 v=1              & 25.4 (7.7) Jy                                     & 105.4 (21.3) Jy kms$^{-1}$  \\
 \hline  

   \end{tabular}
\end{table}

\begin{table}
   \caption{\label{rhya_results}  R Hya results}
   \centering
   \begin{tabular}{lrr}
   \hline\hline 
   & \multicolumn{2}{c}{June 04 2014 (stellar phase $\sim$ 0.55)}\\
 \multicolumn{1}{c}{Line}               &   \multicolumn{1}{c}{Peak (rms)}      &  \multicolumn{1}{c}{Integrated (rms)} \\
 \hline
 CO 4-3                             & 4.77 (0.09) K & 44.9 (0.3)  K kms$^{-1}$\\
 $^{29}$SiO 11-10 v=0     &  0.54 (0.15) K & 1.71 (0.42) K kms$^{-1}$\\
  H$_2$O 437 GHz          & ... (3.5) Jy & \multicolumn{1}{c}{...}\\
 H$_2$O 439 GHz         &  ... (3.8) Jy & \multicolumn{1}{c}{...}\\
 H$_2$O 471 GHz         & ... (4.6) Jy &  \multicolumn{1}{c}{...}\\
 H$_2$O 474 GHz         & ... (5.9) Jy   &  \multicolumn{1}{c}{...}\\
 SiO 11-10 v=1               &   ... (7.1) Jy &  \multicolumn{1}{c}{...}\\
\hline  
   \end{tabular}
\end{table}
   
     
\begin{table}
   \caption{\label{rsvir_results}  RS Vir results}
   \centering
   \begin{tabular}{lrr}
   \hline\hline 
   & \multicolumn{2}{c}{May 09 2013 (stellar phase $\sim$ 0.2)} \\
 \multicolumn{1}{c}{Line}               &   \multicolumn{1}{c}{Peak (rms)}      &  \multicolumn{1}{c}{Integrated (rms)} \\
 \hline
 CO 4-3                        & 1.01 (0.10) K    & 6.20 (0.24) K kms$^{-1}$                                                                    \\
 $^{29}$SiO 11-10 v=0  &  ...    (0.14) K    & \multicolumn{1}{c}{...}                                                                         \\                              
 H$_2$O 437 GHz         &  {\bf 502.3 (7.1)} Jy & 631.0 (16.0) Jy kms$^{-1}$                                                         \\
 H$_2$O 439 GHz         &  21.0 (6.2) Jy   & 55.4 (13.9) Jy kms$^{-1}$                                                                 \\
 H$_2$O 471 GHz         & 17.2 (4.1) Jy    & \multicolumn{1}{c}{...}                                                                        \\
 H$_2$O 474 GHz         & 42.6 (7.3) Jy   & 75.7 (16.5)  Jy kms$^{-1}$                                                                  \\
 SiO 11-10 v=1               & ... (6.8) Jy & \multicolumn{1}{c}{...}    \\
\hline  
& \multicolumn{2}{c}{May 20 2015  (stellar phase $\sim$ 0.2)} \\
 \multicolumn{1}{c}{Line}              &   \multicolumn{1}{c}{Peak (rms)}      &  \multicolumn{1}{c}{Integrated (rms)} \\
 \hline
 CO 4-3                         & 0.72 (0.13) K  & 4.63 (0.29) K kms$^{-1}$ \\
 $^{29}$SiO 11-10 v=0  &    ...     (0.22) K  &    \multicolumn{1}{c}{...}        \\                              
 H$_2$O 437 GHz         & {\bf 56.6 (12.5)} Jy & 111.5 (28.0) Jy kms$^{-1}$ \\
 H$_2$O 439 GHz         & 33.1 (11.5) Jy & \multicolumn{1}{c}{...} \\
 H$_2$O 471 GHz         & 14.4 (6.1) Jy   & \multicolumn{1}{c}{...} \\
 H$_2$O 474 GHz          & 39.5 (11.2) Jy & 72.9 (25.1) kms$^{-1}$ \\
 SiO 11-10 v=1               & ... (9.6) Jy & \multicolumn{1}{c}{...}    \\
\hline  
   \end{tabular}
\end{table}

\begin{table}
   \caption{\label{rleo_results}  R Leo results}
   \centering
   \begin{tabular}{lrr}
   \hline\hline 
   & \multicolumn{2}{c}{May 20 2015  (stellar phase $\sim$ 0.55)} \\
 \multicolumn{1}{c}{Line}               &   \multicolumn{1}{c}{Peak (rms)}      &  \multicolumn{1}{c}{Integrated (rms)} \\
 \hline
 CO 4-3                             & 5.50 (0.32) K  & 50.4 (0.7) K kms$^{-1}$     \\
 $^{29}$SiO 11-10 v=0     &  0.46 K            &  6.17 (1.03) K kms$^{-1}$     \\
  H$_2$O 437 GHz          & ... (12.0) Jy      & \multicolumn{1}{c}{...}   \\
 H$_2$O 439 GHz         &  ... (12.2) Jy       &  \multicolumn{1}{c}{...}   \\
 H$_2$O 471 GHz         & ... (15.3) Jy        & \multicolumn{1}{c}{...}   \\
 H$_2$O 474 GHz         & ... (28.7) Jy        &  \multicolumn{1}{c}{...}   \\
 SiO 11-10 v=1               & ... (28.4) Jy        &  \multicolumn{1}{c}{...}   \\
\hline  
   \end{tabular}
\end{table}

\begin{table}
   \caption{\label{raql_results} R Aql results}
   \centering
   \begin{tabular}{rrr}
   \hline\hline 
   & \multicolumn{2}{c}{Apr 13 2014 (stellar phase $\sim$ 0.25)} \\
 \multicolumn{1}{c}{Line}               &   \multicolumn{1}{c}{Peak (rms)}      &  \multicolumn{1}{c}{Integrated (rms)} \\
 \hline
 CO 4-3                          & 4.13 (0.08) K       & 47.4 (0.2) K kms$^{-1}$                                                          \\
$^{29}$SiO 11-10 v=0  &    ... (0.11) K         &     \multicolumn{1}{c}{...}                                                            \\
H$_2$O 437 GHz         &  {\bf 136.3 (4.1)} Jy     & 228.3 (9.2) Jy kms$^{-1}$                                                   \\
H$_2$O 439 GHz         &  22.6 (5.8) Jy       & 72.9 (13.0) Jy kms$^{-1}$                                                          \\
H$_2$O 471 GHz         &  24.6 (3.7) Jy        & 107.9 (8.3) Jy kms$^{-1}$                                                        \\
H$_2$O 474 GHz         & ... (8.6) Jy             &  \multicolumn{1}{c}{...}                                                               \\
SiO 11-10 v=1               & ... (4.8) Jy             &   \multicolumn{1}{c}{...}                                                               \\
\hline  
 & \multicolumn{2}{c}{May 20 2015 (stellar phase $\sim$ 0.60)} \\
 \multicolumn{1}{c}{Line}               &   \multicolumn{1}{c}{Peak (rms)}      &  \multicolumn{1}{c}{Integrated (rms)} \\
 \hline
 CO 4-3                         & 4.07 (0.11) K                                    & 43.8 (0.3)  K kms$^{-1}$  \\
$^{29}$SiO 11-10 v=0  &   ... (0.18) K                                      &  \multicolumn{1}{c}{...} \\
H$_2$O 437 GHz         & ... (10.7)   Jy                                        & \multicolumn{1}{c}{...} \\
H$_2$O 439 GHz         & ... (12.9)   Jy                                        & \multicolumn{1}{c}{...} \\
H$_2$O 471 GHz         & ... (5.2)     Jy                                        & \multicolumn{1}{c}{...} \\
H$_2$O 474 GHz         & ... (8.6)     Jy                                         & \multicolumn{1}{c}{...} \\
SiO 11-10 v=1               &  ... (8.1)    Jy                                          & \multicolumn{1}{c}{...} \\
\hline  
   \end{tabular}
\end{table}

\begin{table}
   \caption{\label{rdor_results}  R Dor results }
   \centering
   \begin{tabular}{lrr}
   \hline\hline 
      & \multicolumn{2}{c}{May 18 2015 (stellar phase $\sim$ 0.2)} \\
 \multicolumn{1}{c}{Line}               &   \multicolumn{1}{c}{Peak (rms)}      &  \multicolumn{1}{c}{Integrated (rms)} \\
 \hline
 CO 4-3                             &  9.60 (0.13) K    &  98.1 (0.3) K kms$^{-1}$  \\
 $^{29}$SiO 11-10 v=0     & 1.93 (0.20) K      & 15.4 (0.4) K kms$^{-1}$  \\
  H$_2$O 437 GHz          & 28.8 (9.7) Jy        & \multicolumn{1}{c}{...}   \\
   H$_2$O 439 GHz         &  30.7 (11.3) Jy      & \multicolumn{1}{c}{...}   \\
    H$_2$O 471 GHz         & 18.7 (6.5) Jy        & 42.6 (14.5) Jy kms$^{-1}$  \\
     H$_2$O 474 GHz         & 42.9 (10.8) Jy      & \multicolumn{1}{c}{...}   \\
      SiO 11-10 v=1               & ... (13.0) Jy         & \multicolumn{1}{c}{...}   \\
      \hline  
   \end{tabular}
\end{table}

\begin{table}
   \caption{\label{rcrt_results}  R Crt results. It was not possible to determine the stellar phases for these observations.}
   \centering
   \begin{tabular}{lrr}
   \hline\hline 
  & \multicolumn{2}{c}{May 08 2013} \\
   \multicolumn{1}{c}{Line}               &   \multicolumn{1}{c}{Peak (rms)}      &  \multicolumn{1}{c}{Integrated (rms)} \\
  \hline
CO 4-3                              & 2.18 (0.086)  K          &     37.0 (0.2) K kms$^{-1}$ \\
$^{29}$SiO 11-10 v=0              &  ... (0.197) K              &     3.5 (0.6) K kms$^{-1}$ \\
H$_2$O 437 GHz              & ... (14.6) Jy               &     \multicolumn{1}{c}{...} \\
H$_2$O 439  GHz            & ... (12.2)  Jy               &     \multicolumn{1}{c}{...} \\ 
H$_2$O 471  GHz            & ... (5.4) Jy                   &      \multicolumn{1}{c}{...} \\ 
H$_2$O 474  GHz            & ... (11.1) Jy                 &   \multicolumn{1}{c}{...} \\ 
SiO 11-10 v=1                   & ... (10.8) Jy                 &    \multicolumn{1}{c}{...} \\        
\hline
 & \multicolumn{2}{c}{April 12 2014}\\
 \multicolumn{1}{c}{Line}               &   \multicolumn{1}{c}{Peak (rms)}      &  \multicolumn{1}{c}{Integrated (rms)} \\
\hline
 CO 4-3                                            & 2.19 (0.066) K  &    36.6 (0.2) K kms$^{-1}$ \\
$^{29}$SiO 11-10 v=0                             & ... (0.117) K       &    5.4 (0.4) K kms$^{-1}$ \\
H$_2$O 437  GHz                          & ... (4.6) Jy                & \multicolumn{1}{c}{...} \\
H$_2$O 439  GHz                          & ... (9.0) Jy                & \multicolumn{1}{c}{...} \\
H$_2$O 471   GHz                         & ... (3.3) Jy                & \multicolumn{1}{c}{...} \\
H$_2$O 474   GHz                         & ... (4.5) Jy                & \multicolumn{1}{c}{...} \\
SiO 11-10 v=1                                & ... (4.5) Jy                 & \multicolumn{1}{c}{...} \\
 \hline
 & \multicolumn{2}{c}{June 08 2014}\\
 \multicolumn{1}{c}{Line}               &   \multicolumn{1}{c}{Peak (rms)}      &  \multicolumn{1}{c}{Integrated (rms)} \\
 \hline
CO 4-3                                           &     2.13 (0.078) K                              &     36.1 (0.2) K kms$^{-1}$ \\
$^{29}$SiO 11-10 v=0                            &    ... (0.121) K                                   &      4.0 (0.4) K kms$^{-1}$ \\
H$_2$O 437 GHz                           &   ... (3.0) Jy                                      & \multicolumn{1}{c}{...} \\
H$_2$O 439 GHz                           &  ... (3.8)  Jy                                      & \multicolumn{1}{c}{...} \\
H$_2$O 471 GHz                           &  ... (3.8) Jy                                       & \multicolumn{1}{c}{...} \\
H$_2$O 474 GHz                           &  ... (6.1) Jy                                       & \multicolumn{1}{c}{...} \\
SiO 11-10 v=1                                &  ...  (6.1) Jy                                       & \multicolumn{1}{c}{...} \\
 \hline
  & \multicolumn{2}{c}{June 15 2014}\\
 \multicolumn{1}{c}{Line}               &   \multicolumn{1}{c}{Peak (rms)}      &  \multicolumn{1}{c}{Integrated (rms)} \\
 \hline
CO 4-3                                           &  1.90 (0.117) K                                 & 31.3 (0.4)  K kms$^{-1}$ \\
$^{29}$SiO 11-10 v=0                    &  ... (0.194) K                                     & 3.7 (0.5) K kms$^{-1}$ \\
H$_2$O 437 GHz                          & ... (6.9) Jy                                         &  \multicolumn{1}{c}{...} \\
H$_2$O 439 GHz                           & ... (7.7) Jy                                        &   \multicolumn{1}{c}{...} \\
H$_2$O 471 GHz                           & ... (5.5) Jy                                        & \multicolumn{1}{c}{...} \\
H$_2$O 474 GHz                           & ... (11.2) Jy                                      &  \multicolumn{1}{c}{...} \\
SiO 11-10 v=1                                & ... (11.6) Jy                                      &  \multicolumn{1}{c}{...} \\
 \hline
 & \multicolumn{2}{c}{May 20 2015}\\
 \multicolumn{1}{c}{Line}               &   \multicolumn{1}{c}{Peak (rms)}      &  \multicolumn{1}{c}{Integrated (rms)} \\
 \hline
CO 4-3                                          & 2.26 (0.130) K                                   & 34.3 (0.4) K kms$^{-1}$ \\
$^{29}$SiO 11-10 v=0                   & ... (0.208) K                                       & 6.3 (0.6) K   kms$^{-1}$ \\
H$_2$O 437 GHz                         & ... (13.5) Jy                                        &   \multicolumn{1}{c}{...} \\ 
H$_2$O 439 GHz                         & ... (14.1) Jy                                        &  \multicolumn{1}{c}{...} \\ 
H$_2$O 471 GHz                         & ... (6.5) Jy                                          &  \multicolumn{1}{c}{...} \\ 
H$_2$O 474 GHz                         & ... (11.7) Jy                                        &  \multicolumn{1}{c}{...} \\ 
SiO 11-10 v=1                          & ... (10.6) Jy                                        &   \multicolumn{1}{c}{...} \\
   \hline  
   \end{tabular}
   \end{table}
   
\begin{table}
   \caption{\label{rtvir_results}  RT Vir results}
   \centering
   \begin{tabular}{lrr}
   \hline\hline 
  & \multicolumn{2}{c}{April 12 2014 (stellar phase $\sim$ 0.95) } \\
   \multicolumn{1}{c}{Line}               &   \multicolumn{1}{c}{Peak (rms)}      &  \multicolumn{1}{c}{Integrated (rms)} \\
  \hline
CO 4-3                                & 1.29 (0.07) K  & 14.4 (0.2) K  kms$^{-1}$ \\
$^{29}$SiO 11-10 v=0        & 0.49 (0.12) K   & 3.21 (0.32) K  kms$^{-1}$ \\
H$_2$O 437 GHz               & ... (3.8) Jy       &  \multicolumn{1}{c}{...} \\ 
H$_2$O 439 GHz               & ... (5.3) Jy       &  \multicolumn{1}{c}{...} \\ 
H$_2$O 471 GHz               & ...  (3.5) Jy      & \multicolumn{1}{c}{...} \\ 
H$_2$O 474 GHz               & ... (4.6) Jy       & \multicolumn{1}{c}{...} \\ 
SiO 11-10 v=1                & ... (4.1) Jy       & \multicolumn{1}{c}{...} \\
   \hline\hline 
  & \multicolumn{2}{c}{May 20 2015 (stellar phase $\sim$ 0.30)} \\
   \multicolumn{1}{c}{Line}               &   \multicolumn{1}{c}{Peak (rms)}      &  \multicolumn{1}{c}{Integrated (rms)} \\
  \hline
CO 4-3                             &  1.77  (0.13) K   & 17.7 (0.3) K  kms$^{-1}$ \\
$^{29}$SiO 11-10 v=0  &  0.75  (0.22) K   & 2.80 (0.56) K kms$^{-1}$ \\
H$_2$O 437 GHz             & ... (12.5) Jy       & \multicolumn{1}{c}{...} \\
H$_2$O 439 GHz             & ... (11.9) Jy        & \multicolumn{1}{c}{...} \\
H$_2$O 471 GHz             & ... (6.6) Jy         &  \multicolumn{1}{c}{...} \\
H$_2$O 474 GHz             & ... (11.4) Jy       & \multicolumn{1}{c}{...} \\
SiO 11-10 v=1              & ... (10.0) Jy       & \multicolumn{1}{c}{...} \\
\hline  
   \end{tabular}
\end{table}

\section{Mass loss rate determination}
\label{app:CO}
     
\begin{figure*}
  \centering
    \includegraphics[width=7cm]{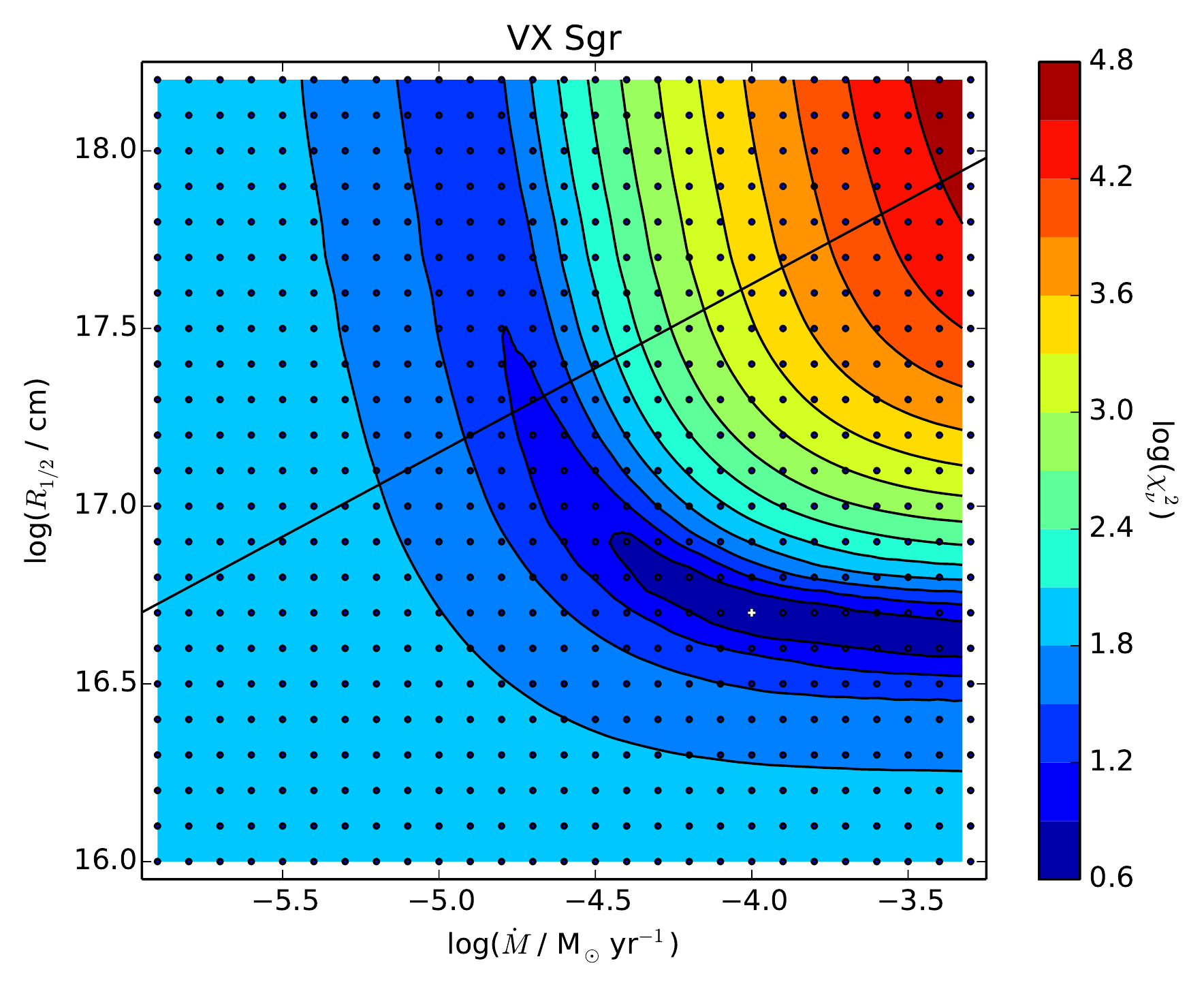}
    \includegraphics[width=7cm]{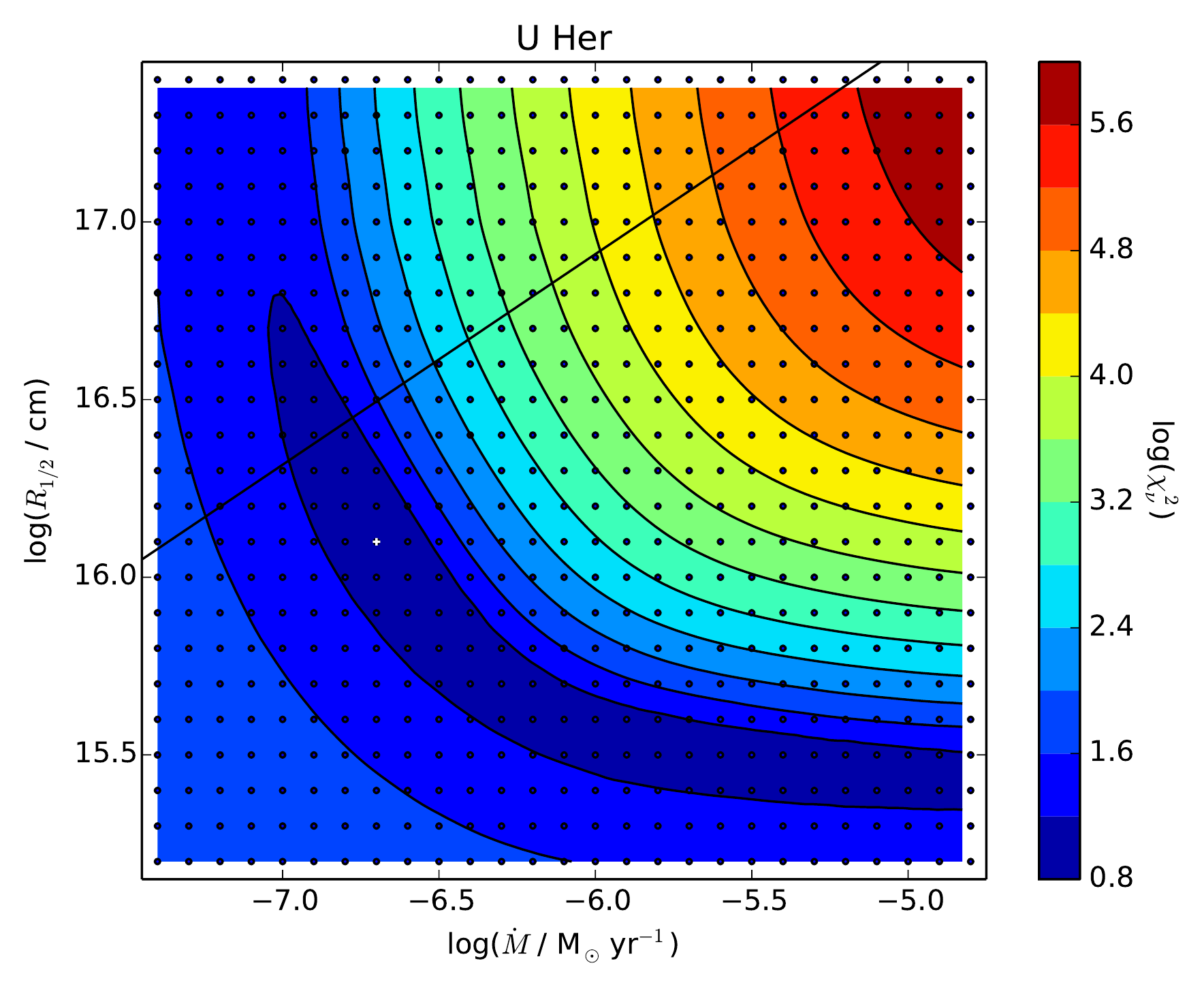}
    \includegraphics[width=7cm]{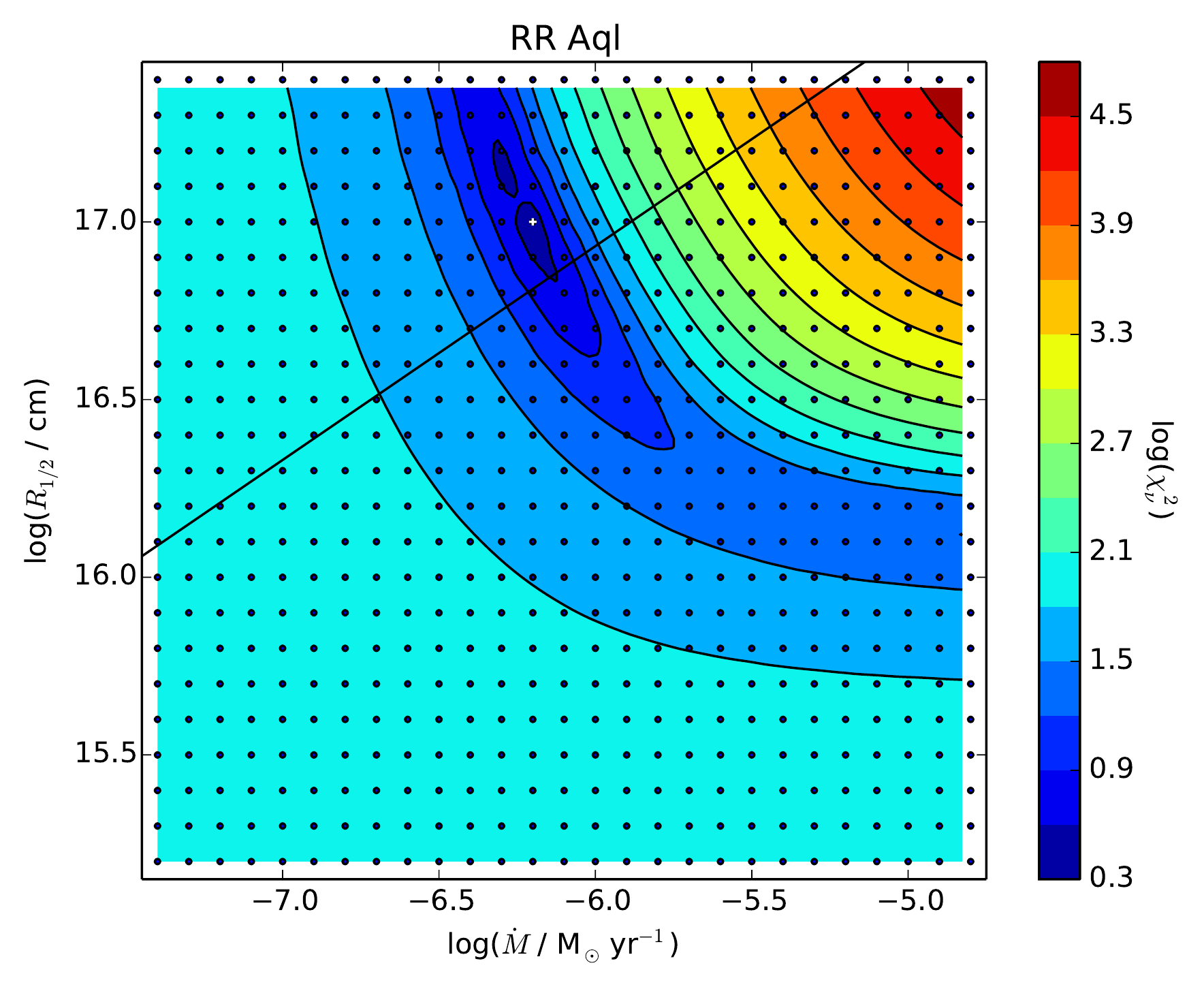}
    \includegraphics[width=7cm]{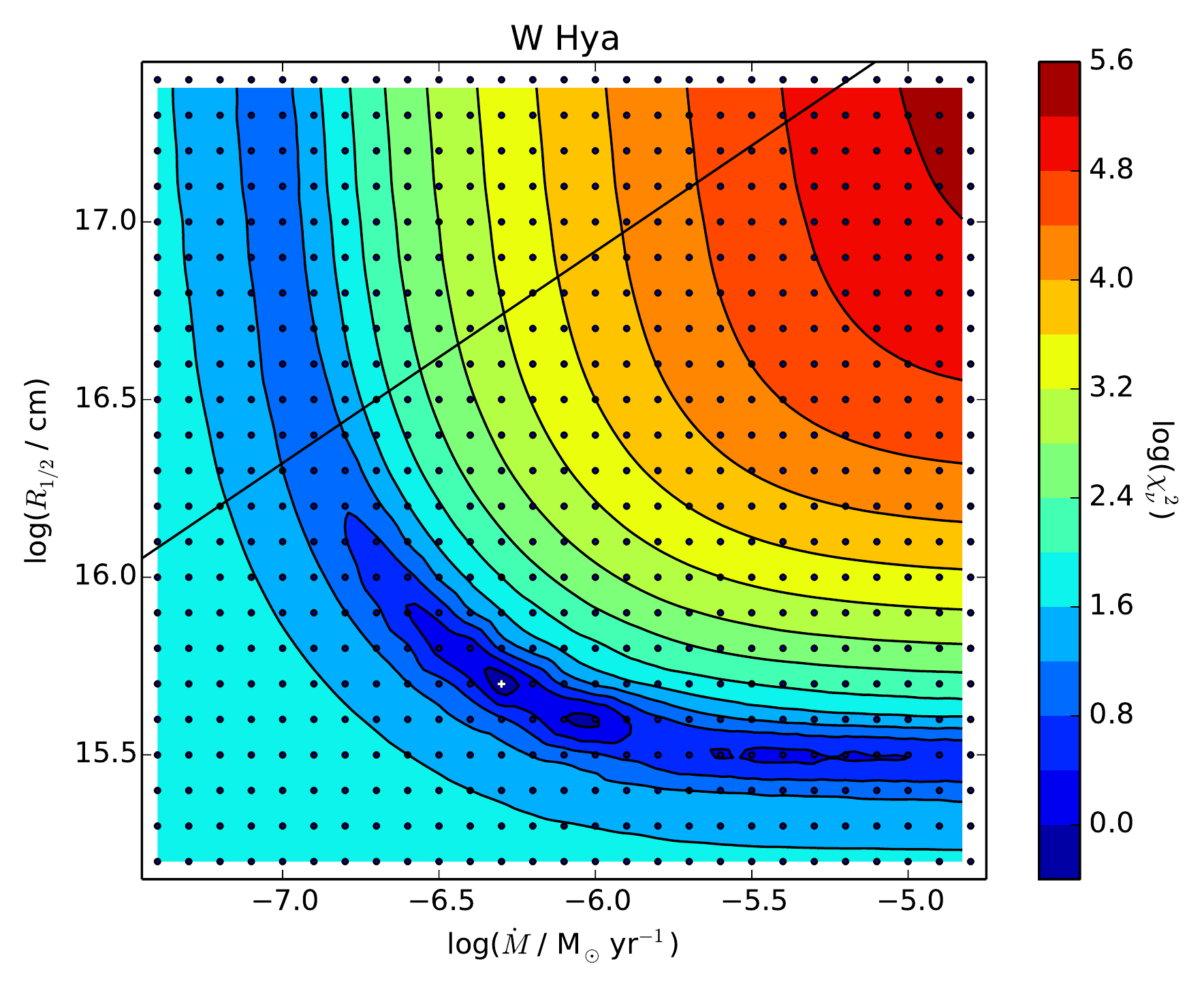}
  \caption{Grid of mass loss rate modelling for VX Sgr, U Her, RR Aql, and W Hya. The contours represent
    the base-10 logarithm of the reduced $\chi^2_\nu$-value from the fitting of CO lines. The horizontal
    axis describes the mass loss rate variation and along the vertical axis the the CO photo-dissociation
    radius ($R_{1/2}$) is varied. The black dots indicate the grid values and the white cross
    indicates the best fit values.
    The solid line represents the CO photo-dissociation radius vs. mass loss dependence as predicted by
    \citet{Mamon1988}.}
              \label{vxsgr_cogrid}
\end{figure*}

\begin{figure*}
  \centering
    \includegraphics[width=7cm]{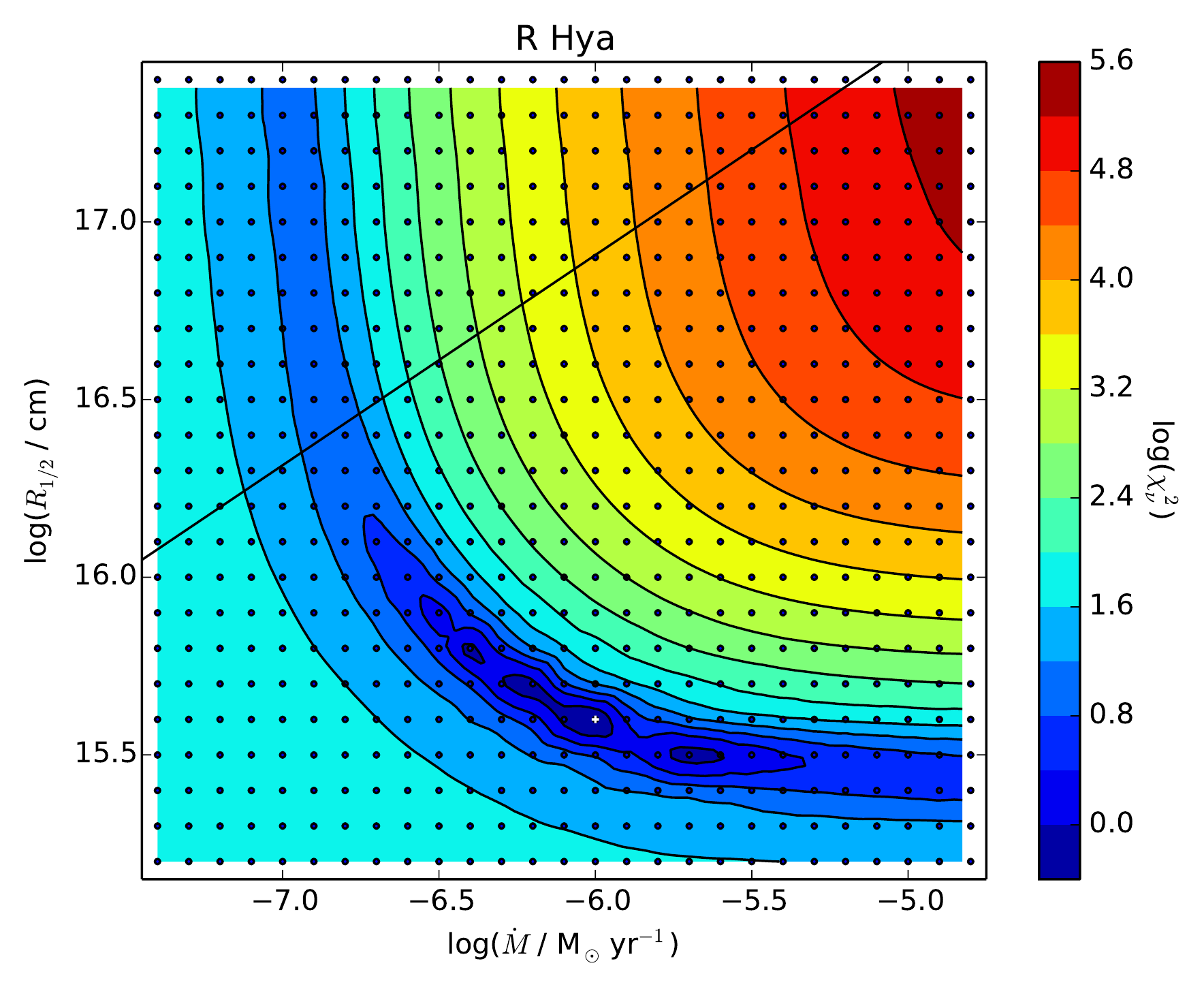}
    \includegraphics[width=7cm]{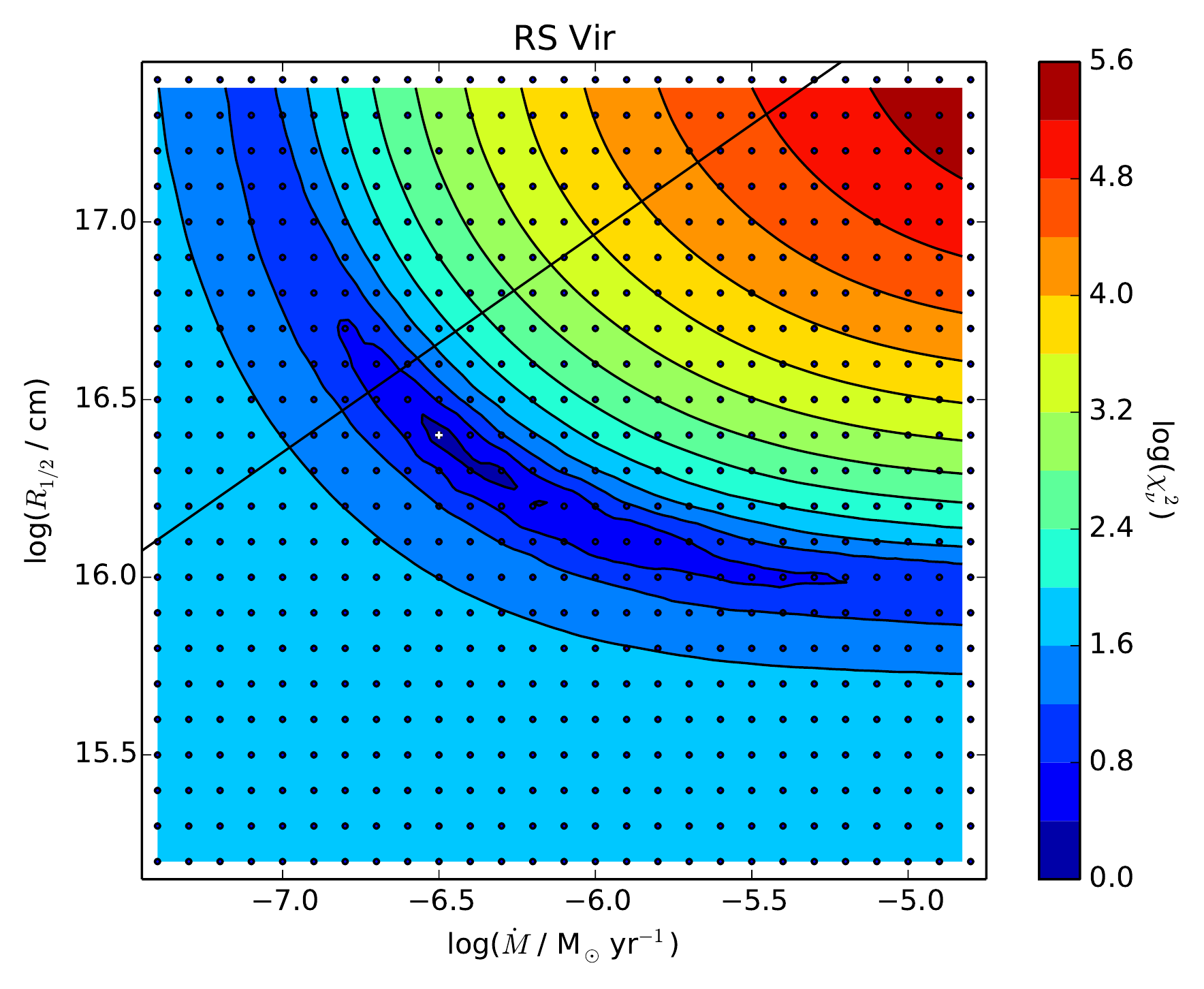}
    \includegraphics[width=7cm]{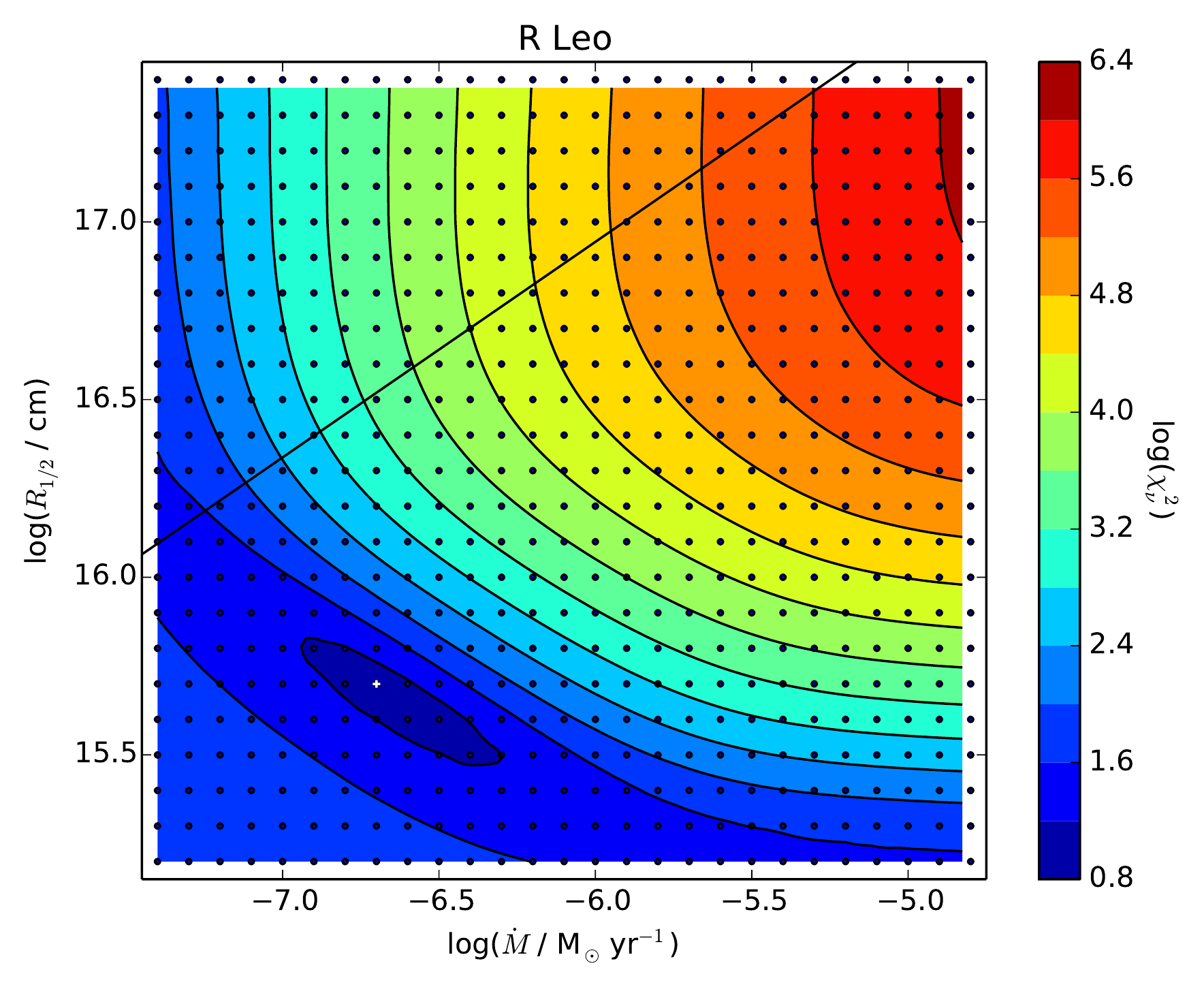}
    \includegraphics[width=7cm]{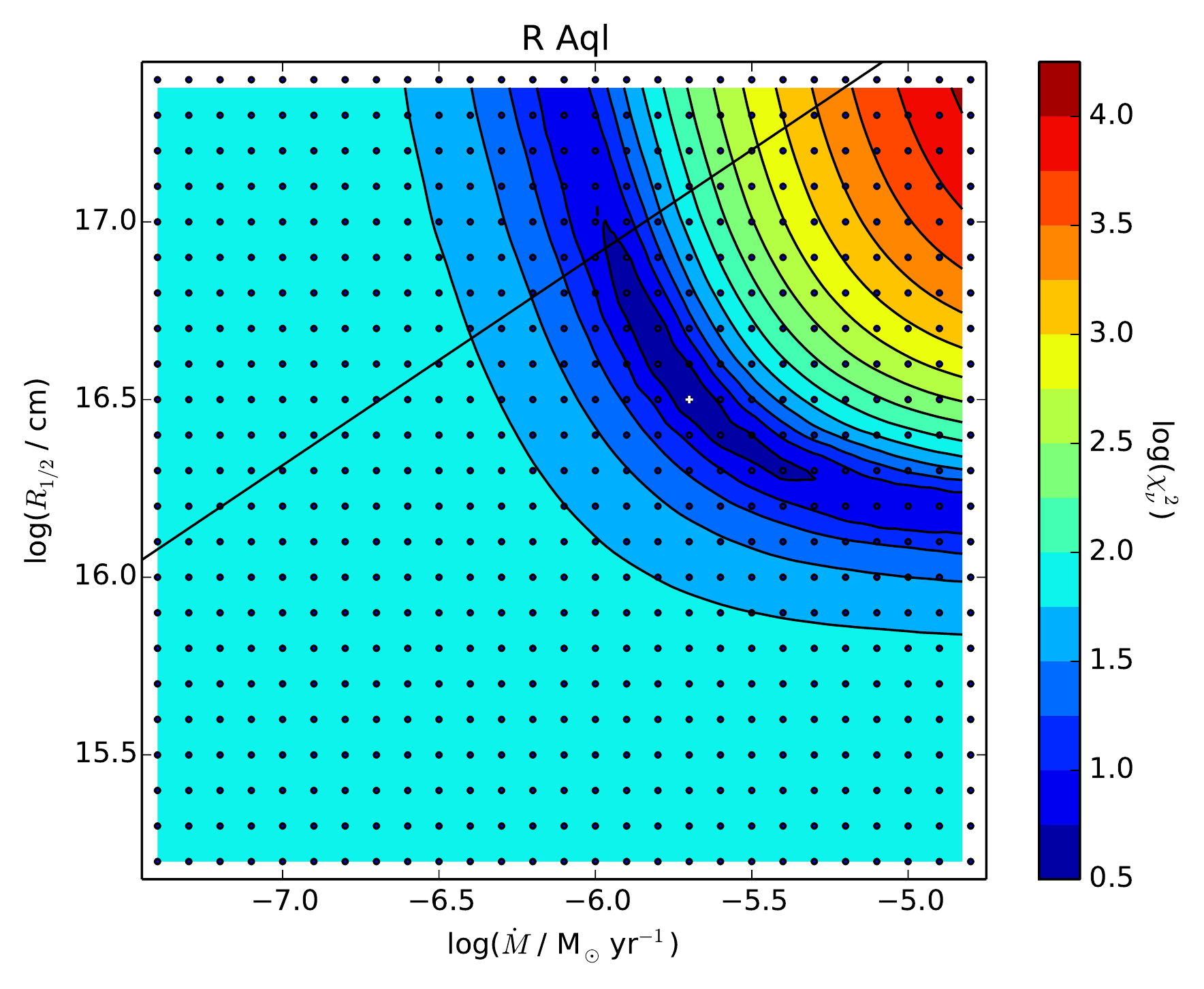}
 \caption{Grid of mass loss rate modelling for R Hya, RS Vir, R Leo, and R Aql. Scales are as in Fig.~\ref{vxsgr_cogrid}}
              \label{rhya_cogrid}
\end{figure*}

\begin{figure*}
\centering
  \includegraphics[width=7cm]{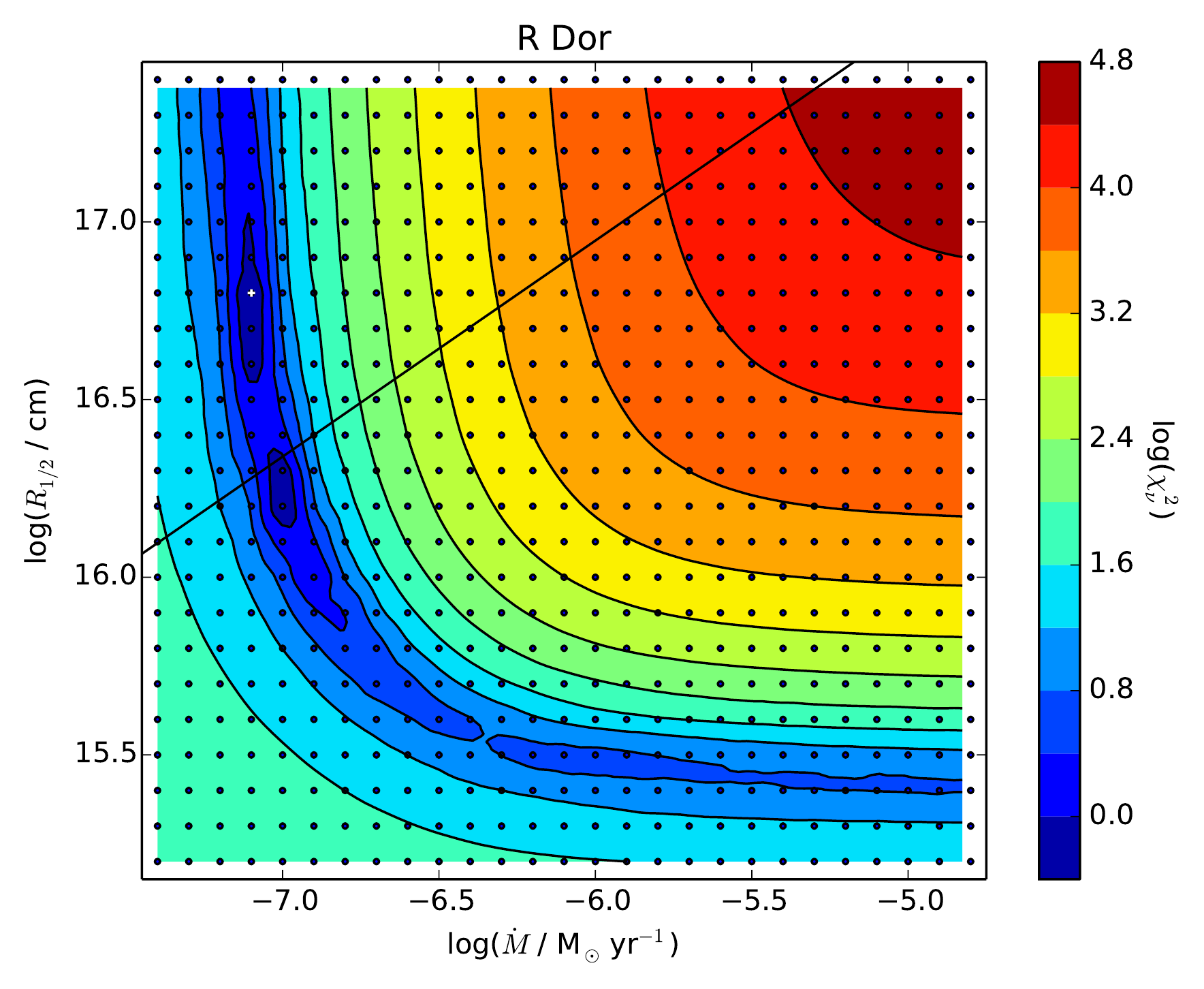}
  \includegraphics[width=7cm]{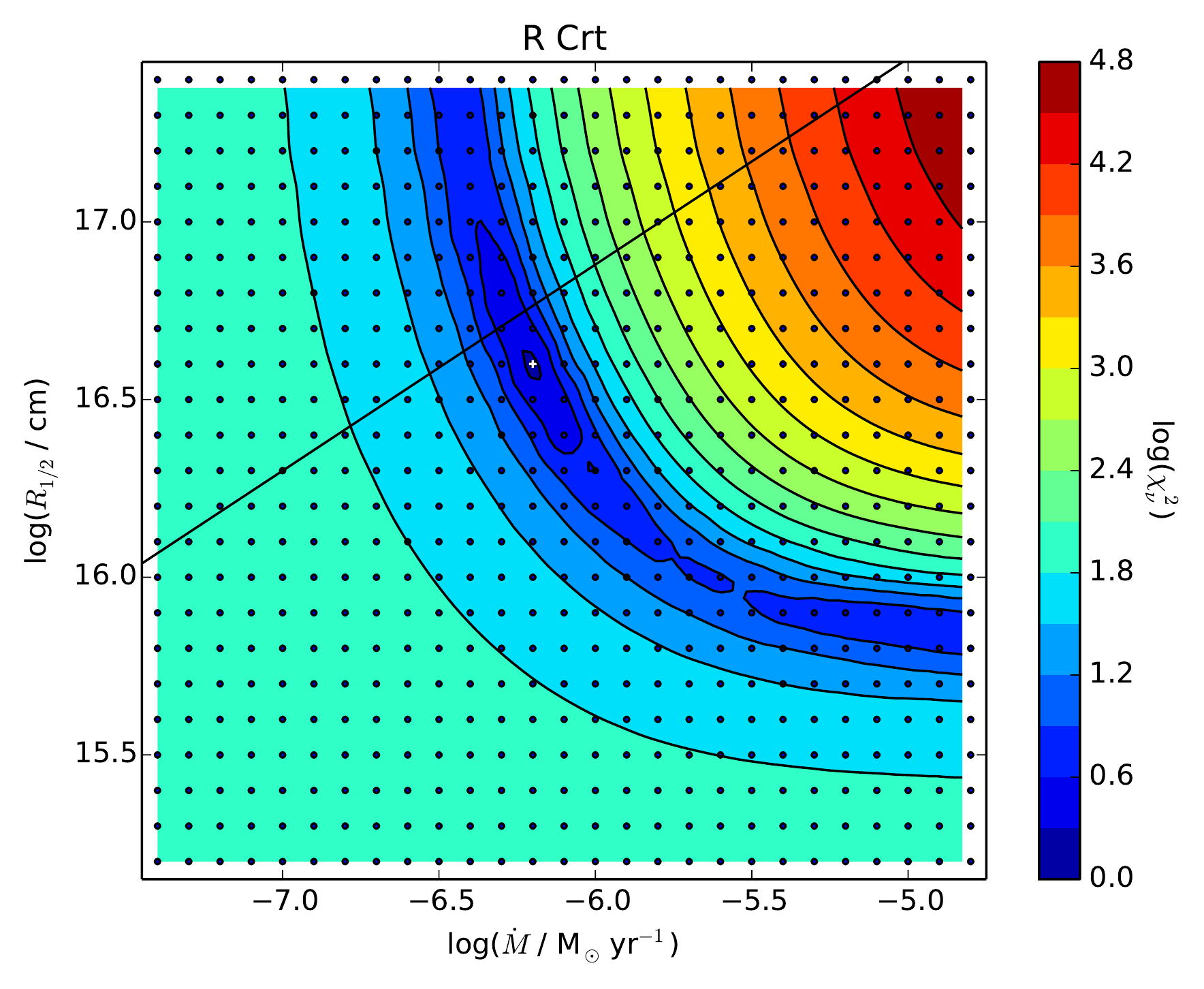}
  \includegraphics[width=7cm]{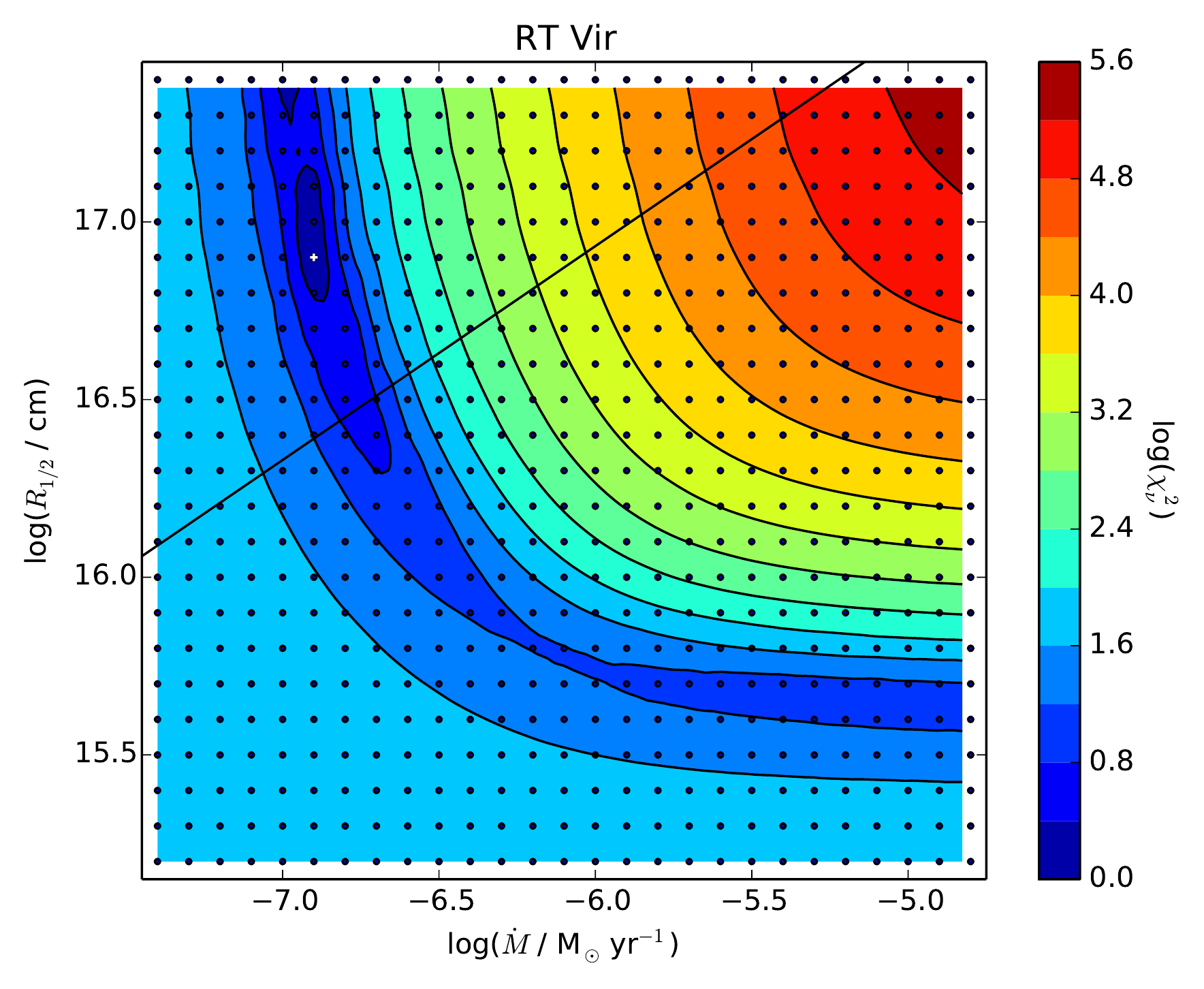}
 \caption{Grid of mass loss rate modelling for R Dor, R Crt, and RT Vir. Scales are as in Fig.~\ref{vxsgr_cogrid}}
              \label{rdor_cogrid}
\end{figure*}

We also performed CO line modelling using the CO(4-3) data in this paper, as well as additional CO line data
listed in Table~\ref{co_lines} to determine the stellar mass loss rates.
We use the same ALI code as used for our water modelling in Sect.~\ref{sect:ali}. For CO and H$_2$
collisions we use the coefficients calculated by \citet{Yang2010}.
We adopt a density radial model as given by the mass loss rate and a constant expansion velocity as determined
from the width of our observed CO 4-3 lines, and a kinetic temperature profile as outlined in Sect.~\ref{sect:smooth-wind}.
The CO abundance used is $5 \times 10^{-4}$ near the centre and is half that at $r = R_{1/2}$ and declines exponentially
out to the cloud radius at about $R=R_{1/2}/0.6$ where the envelope is truncated.
Here we construct a grid by varying mass loss rate ${\dot{M}}$
and CO radius $R_{1/2}$ and display our fitting results by means of a reduced $\chi^2_\nu$ value calculated from
observed line intensities and modelled ones using appropriate beam sizes. For each source, between three and seven
integrated line intensities are used (see Table~\ref{co_lines}). In the case where the CO 4-3 line was observed at different epochs,
the integrated intensity obtained at the date with smallest uncertainty was used. The grid results
for our targeted stars are displayed in Fig.~\ref{vxsgr_cogrid} to Fig.~\ref{rdor_cogrid}. In each of the figures,
a line is shown corresponding to the relation between the mass loss rate and the CO photo-dissociation radius (and expansion velocity)
found by \citet{Mamon1988}. The modelling results are summarised in Table~\ref{co_modelling}.

\begin{table*}
          \caption{\label{co_lines} CO line intensities, in $\mathrm{K\, km\, s^{-1}}$, used in the modelling}
\centering
   \begin{tabular}{lcccccc}
   \hline\hline
Star      & $J=1-0$                 &  $J=2-1$            & $J=3-2$    & $J=4-3$   &  $J=6-5$       &        $N$      \\
\hline
VX Sgr & ---                         & 32.5\tablefoottext{a}  & 94.5\tablefoottext{a}  & 99.7\tablefoottext{b}  & ---& 3 \\
U Her  & 2.1\tablefoottext{c}& ---                              & 7.3\tablefoottext{d}    & 8.0\tablefoottext{b},4.9\tablefoottext{d}      & ---& 4 \\
RR Aql & 2.9\tablefoottext{e}&11.8\tablefoottext{f}   & 13.7\tablefoottext{f}  & 11.7\tablefoottext{b}  & ---& 4 \\
W Hya  & ---               &  12.1\tablefoottext{f}            &       36.5\tablefoottext{f}& 52.7\tablefoottext{b}  & ---& 3 \\
R Hya  & ---      & 9.7\tablefoottext{f},17.0\tablefoottext{c} & 30.2\tablefoottext{f},41.5\tablefoottext{a}& 44.9 \tablefoottext{b} & 51.7\tablefoottext{a} & 6 \\
RS Vir & 2.3\tablefoottext{e}     &19.5\tablefoottext{e}             & --- & 6.2\tablefoottext{b}  & --- & 3 \\
R Leo  & 2.4\tablefoottext{i}, 4.1\tablefoottext{h} & 14.9\tablefoottext{f}           & 40.0\tablefoottext{f} & 50.4\tablefoottext{b}  & --- & 5 \\
R Aql  & ---                  & 19.9\tablefoottext{f} &  35.6\tablefoottext{f},45.0\tablefoottext{d} & 47.4\tablefoottext{b},36.9\tablefoottext{d} & --- & 5 \\
R Dor  &  --                  & 37.5\tablefoottext{f} & 64.1\tablefoottext{f} & 98.1\tablefoottext{b}  & 120\tablefoottext{j} & 4 \\
R Crt  & 5.0\tablefoottext{g} & 22.0\tablefoottext{f},24.0\tablefoottext{g} & 26.7\tablefoottext{f},39.2\tablefoottext{g} & 37.0\tablefoottext{b},54.1\tablefoottext{g}& --- & 7 \\
RT Vir & 5.2\tablefoottext{g} & 11.8\tablefoottext{f},14.1\tablefoottext{g} & 13.5\tablefoottext{f},19.0\tablefoottext{g} & 14.4\tablefoottext{b}, 16.1\tablefoottext{g}  & --- & 7 \\
 \hline
           \end{tabular}
\tablefoot{
\tablefoottext{a}{\citet{DeBeck2010}.}
\tablefoottext{b}{$J=4-3$ values determined in this work.}
\tablefoottext{c}{\citet{Josselin1998}.}
\tablefoottext{d}{\citet{Young1995}.}
\tablefoottext{e}{\citet{Groenewegen1999}.}
\tablefoottext{f}{APEX pointing data: \url{http://www.apex-telescope.org/observing/pointing/spectra}.}
\tablefoottext{g}{\citet{Kerschbaum1999}.}
\tablefoottext{h}{\citet{Teyssier2006}.}
\tablefoottext{i}{\citet{Olofsson1998}.}
\tablefoottext{j}{Commissioning data for APEX Band 9 (unpublished).}
          }
\end{table*}

\begin{table}
          \caption{\label{co_modelling} CO modelling results}
\centering
   \begin{tabular}{lrcccc}
   \hline\hline
Star      &               & \multicolumn{2}{c}{Free fit} & \multicolumn{2}{c}{Fixed size\tablefoottext{a}} \\
          & \multicolumn{1}{c}{$v_\mathrm{e}$}            & $R_{1/2}$       &   $\dot{M}$         & $R_{1/2}$       &   $\dot{M}$                \\
          & \multicolumn{1}{c}{($\mathrm{km\, s^{-1}}$)}  & (cm)            &  (M$_{\sun}\,$yr$^{-1}$)  & (cm)            &  (M$_{\sun}\,$yr$^{-1}$)      \\
\hline
VX Sgr &   22.0   & 5.0 10$^{16}$    & 1.0 10$^{-4}$  & 2.0 10$^{17}$    & 2.0 10$^{-5}$  \\
U Her  &    8.8   & 1.3 10$^{16}$    & 2.0 10$^{-7}$  & 2.5 10$^{16}$    & 1.3 10$^{-7}$  \\
RR Aql &    7.0   & 1.0 10$^{17}$    & 6.3 10$^{-7}$  & 7.1 10$^{16}$    & 5.6 10$^{-7}$  \\
W Hya  &    8.2   & 5.0 10$^{15}$    & 5.0 10$^{-7}$  & 2.5 10$^{16}$    & 1.3 10$^{-7}$  \\
R Hya  &    9.0   & 4.0 10$^{15}$    & 1.0 10$^{-6}$  & 2.5 10$^{16}$    & 1.4 10$^{-7}$  \\
RS Vir &    4.5   & 2.5 10$^{16}$    & 3.2 10$^{-7}$  & 3.5 10$^{16}$    & 2.2 10$^{-7}$  \\
R Leo  &    6.5   & 5.0 10$^{15}$    & 2.0 10$^{-7}$  & 1.6 10$^{16}$    & 3.2 10$^{-8}$  \\
R Aql  &    9.0   & 3.2 10$^{16}$    & 2.0 10$^{-6}$  & 8.9 10$^{16}$    & 8.9 10$^{-7}$  \\
R Dor  &    5.7   & 6.3 10$^{16}$    & 7.9 10$^{-8}$  & 2.0 10$^{16}$    & 8.9 10$^{-8}$  \\
R Crt  &   11.0   & 4.0 10$^{16}$    & 6.3 10$^{-7}$  & 6.8 10$^{16}$    & 4.5 10$^{-7}$  \\
RT Vir &    7.0   & 7.9 10$^{16}$    & 1.3 10$^{-7}$  & 3.0 10$^{16}$    & 1.8 10$^{-7}$  \\
 \hline
    \end{tabular}
  \tablefoot{
            \tablefoottext{a}{Setting size according to \citet{Mamon1988}}
}
\end{table}

In all of the figures, the region where the models are in closest agreement with the observations (dark blue areas) show
a similar shape for all our sources. Reasonable fits can often occur for a range of mass loss rates, but only when adopting the
corresponding size. The fitting is done by comparing integrated line intensities and not by fitting the shape of the line profiles.
The latter would have been preferable but we do not have access to the spectral information of most other CO lines
than those presented here. Typically there is a difference in line profiles over the range of acceptable combinations
of $\dot{M}$ and $R_{1/2}$. For larger mass loss rates and smaller $R_{1/2}$, the CO 4-3 profile is parabolic while
for lower mass loss rates the shape starts to become flat-topped and even double-peaked. This is a resolution effect in that
for $\dot{M}$ and large $R_{1/2}$, the CSE emission is more likely to be resolved by the adopted beam size \citep{Olofsson1982}.
The \citet{Mamon1988} relation tends to agree better with the results of the stars with lower mass loss rates
($\dot{M} \la 10^{-6}\,\mathrm{M_\sun\, yr^{-1}}$). The best-fit results of $\dot{M}$ and $R_{1/2}$
are shown in Table~\ref{co_modelling}, along with best-fit values corresponding to the \citet{Mamon1988} size--mass
loss rate dependence (the position along the line in Fig.~\ref{vxsgr_cogrid} to Fig.~\ref{rdor_cogrid} where it
crosses the dark blue area). For most of our sources (8 out of 11), the CO size determined by the free fit is smaller than that determined by
the \citet{Mamon1988} relation. This is in agreement with the recent work by \citet{Saberi2019} who suggest that the CO size due to
photodissociation (and including shielding effects) is smaller that the size reported by \citet{Mamon1988}.
The determined mass loss rates are in reasonable agreement with earlier estimates \citep[e.g.][]{DeBeck2010}.


\end{appendix}

\end{document}